\documentclass[twocolumn,tighten]{aastex62}

\usepackage{enumitem}

\usepackage{amsmath}
\usepackage{subfigure}
\usepackage{graphicx}

\usepackage{ascii}
\usepackage[T1]{fontenc}

\begin{document}

\title{A Dramatic Decrease in Carbon Star Formation in M31}

\author[0000-0003-4850-9589]{M.~L.\ Boyer}
\affiliation{STScI, 3700 San Martin Drive, Baltimore, MD 21218 USA}

\correspondingauthor{M.~L.\ Boyer}
\email{mboyer@stsci.edu}

\author[0000-0002-7502-0597]{B.~F. Williams} 
\affiliation{Department of Astronomy, Box 351580, University of Washington, Seattle, WA 98195, USA}

\author{B. Aringer}
\affiliation{Department of Physics and Astronomy Galileo Galilei, University of Padova, Vicolo dell'Osservatorio 3, I-35122 Padova, Italy}
\affiliation{Osservatorio Astronomico di Padova -- INAF, Vicolo dell'Osservatorio 5, I-35122 Padova, Italy}

\author[0000-0002-3759-1487]{Y. Chen}
\affiliation{Department of Physics and Astronomy Galileo Galilei, University of Padova, Vicolo dell'Osservatorio 3, I-35122 Padova, Italy}

\author[0000-0002-1264-2006]{J.~J. Dalcanton}
\affiliation{Department of Asstronomy, Box 351580, University of Washington, Seattle, WA 98195, USA}

\author[0000-0002-6301-3269]{L. Girardi}
\affiliation{Osservatorio Astronomico di Padova -- INAF, Vicolo dell'Osservatorio 5, I-35122 Padova, Italy}

\author[0000-0001-8867-4234]{P. Guhathakurta}
\affiliation{University of California Observatories/Lick Observatory, University of California, 1156 High St., Santa Cruz, CA 95064}

\author[0000-0002-9137-0773]{P. Marigo}
\affiliation{Department of Physics and Astronomy Galileo Galilei, University of Padova, Vicolo dell'Osservatorio 3, I-35122 Padova, Italy}

\author[0000-0002-7134-8296]{K.~A.~G. Olsen}
\affiliation{National Optical Astronomy Observatory, 950 North Cherry Avenue, Tucson, AZ 85719, USA}

\author[0000-0001-9306-6049]{P. Rosenfield}
\affiliation{Eureka Scientific, Inc., 2452 Delmer Street, Oakland CA 94602, USA}

\author[0000-0002-6442-6030]{D.~R. Weisz} 
\affiliation{Department of Astronomy, University of California Berkeley, Berkeley, CA 94720, USA}

\begin{abstract}

We analyze resolved stellar near-infrared photometry of 21 HST fields
in M31 to constrain the impact of metallicity on the formation of
carbon stars. Observations of nearby galaxies show that the carbon
stars are increasingly rare at higher metallicity. Models indicate
that carbon star formation efficiency drops due to the decrease in
dredge-up efficiency in metal-rich thermally-pulsing Asymptotic Giant
Branch (TP-AGB) stars, coupled to a higher initial abundance of
oxygen. However, while models predict a metallicity ceiling above
which carbon stars cannot form, previous observations have not yet
pinpointed this limit. Our new observations reliably separate carbon
stars from M-type TP-AGB stars across 2.6--13.7~kpc of M31's
metal-rich disk using HST WFC3/IR medium-band filters. We find that
the ratio of C to M stars (C/M) decreases more rapidly than
extrapolations of observations in more metal-poor galaxies, resulting
in a C/M that is too low by more than a factor of 10 in the innermost
fields and indicating a dramatic decline in C star formation
efficiency at metallicities higher than ${\rm
  [M/H]}\approx-0.1$~dex. The metallicity ceiling remains undetected,
but must occur at metallicities higher than what is measured in M31's
inner disk (${\rm [M/H]}\gtrsim+0.06$~dex).

\end{abstract}


\section{INTRODUCTION}
\label{sec:intro}

Stars with initial masses of $\approx$0.8--8~$M_\odot$ will pass
through the thermally-pulsing Asymptotic Giant Branch (TP-AGB) phase
at the end of their evolution. The TP-AGB phase not only determines
the subsequent evolution of the star (primarily due to strong mass
loss), but also has a considerable impact on the evolution and
appearance of the host galaxy. TP-AGB stars are a significant
contributor to galaxy chemical enrichment of light elements and heavy
s-processes elements
\citep{KarakasLattanzio2014,Fishlock+2014,Cristallo+2015} and to a
galaxy's dust budget \citep{Matsuura+2009, Boyer+2012, Zhukovska+2013,
  Schneider+2014, Srinivasan+2016}. TP-AGB stars are also among the
brightest objects in galaxies, especially in the near-infrared, and
thus contribute up to 70\% of a galaxy's total integrated light
\citep{Melbourne+2012,MelbourneBoyer2013}. The physics of TP-AGB stars
therefore has major implications for galaxy properties derived from
rest-frame near-infrared light \citep[e.g.,][]{Conroy+2009,
  Johnson+2013, Baldwin+2017}.

Unfortunately, TP-AGB stars are simultaneously important and
notoriously difficult to model, due to complex processes such as
pulsation (both dynamic and thermal), mass loss, convection, and
dredge up. Adding to the difficulty is a dearth of observational
constraints for AGB models. The paucity of usable observations is
rooted in 1) small populations and small number statistics due to
their short lifetime and 2) the historic difficulty in obtaining
infrared observations of individual stars in star-forming galaxies
in and beyond the Local Group.

\begin{figure*}
  \begin{center}
  \includegraphics[width=1\textwidth]{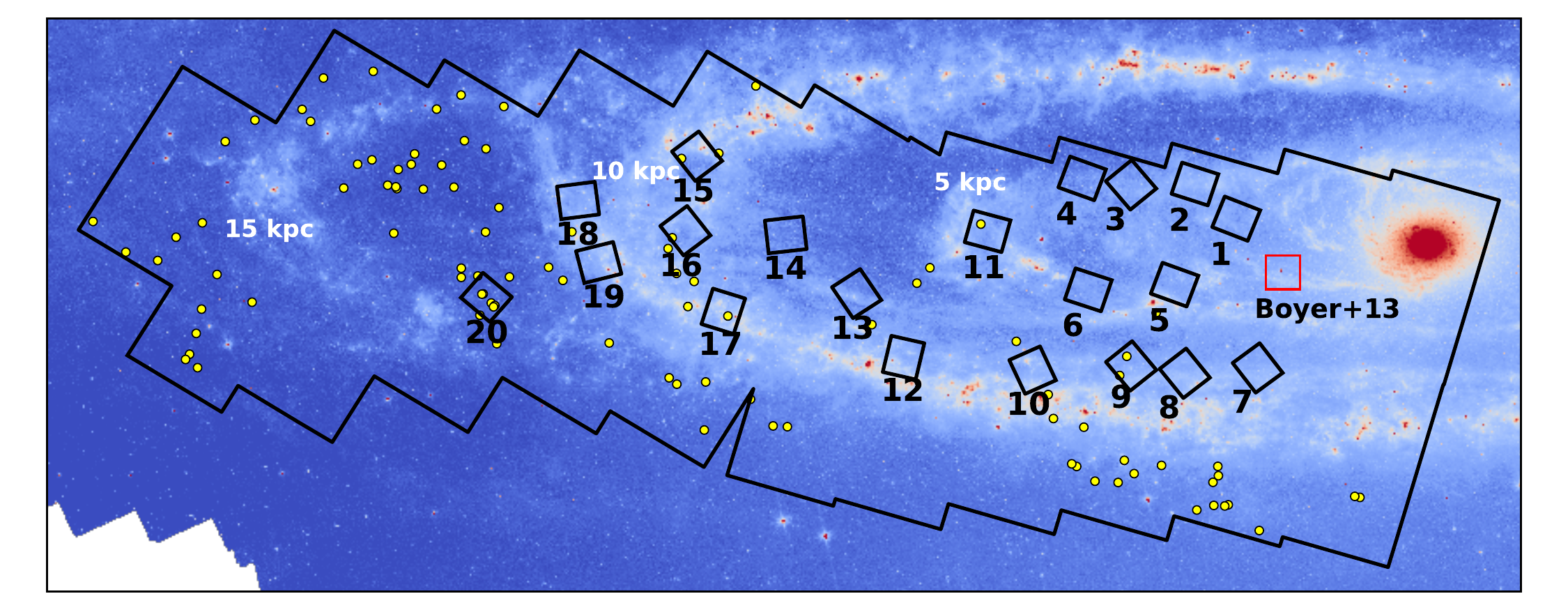}
  \caption{Spitzer 8-\micron\ image of M31 showing the PHAT footprint
    (outer black outline), the pilot field (red; B13), and the 20
    fields observed here (136\arcsec $\times$ 123\arcsec). Yellow dots are spectroscopically-confirmed
    carbon stars from SPLASH \citep{Hamren+2015}. The approximate location of the 5, 10, and 15~kpc star-forming rings are marked for reference. \label{fig:map}}
  \end{center}
\end{figure*}

Recent work has provided new advances in describing the TP-AGB phase
in metal-poor environments \citep{VenturaMarigo2010,
  Marigo+2013,Rosenfield+2014,Rosenfield+2016,Pastorelli+2019,Nanni+2016,
  Karakas+2018}, but metal-rich models remain poorly constrained. In
the Milky Way, TP-AGB stars have intrinsically uncertain Gaia
distances due to the motion of large convection cells
\citep{Chiavassa+2011, Chiavassa+2018}, making their infrared
luminosities unreliable. This limitation makes it impossible to use
Galactic TP-AGB stars to constrain high metallicity models. The next
most nearby metal-rich environment is the Andromeda Galaxy (M31),
which has recently been observed with high precision in the
near-infrared by the Panchromatic Hubble Andromeda Treasury (PHAT)
program \citep{Dalcanton+2012b}. The PHAT program provides a large
population of easily observable, high metallicity TP-AGB stars at a
common distance.

In this work, we target several fields in the region of M31 observed
by the PHAT program with observations designed specifically to
identify the two main TP-AGB stellar types: carbon-rich (C type) and
oxygen-rich (M type).  Separating these two stellar types is key for
calibrating the third dredge up process, which critically determines
when a star becomes carbon rich and hence affects the onset of the
final phases of enhanced mass loss. The formation of C stars is
favored at low metallicities because (1) less oxygen is available to
bind dredged-up carbon into the CO molecule, and (2) the depth of the
third dredge up events increases at low metallicity
\citep{Karakas+2002}. This varying efficiency directly reflects in the
ratio of C to M stars (C/M), which is observed to be anticorrelated
with metallicity. At high metallicity, theoretical models indicate
that the onset of the superwind phase terminates the TP-AGB phase
before the star goes through enough dredge-up events to become C-rich
\citep{WeissFerguson2009,Marigo+2013}, resulting in a predicted
metallicity ceiling for carbon star formation. The COLIBRI models
\citep{Marigo+2013}, for example, predict that this metallicity
ceiling occurs somewhere between ${\rm [M/H]} = +0.1$ and $+0.4$
\citep{Boyer+2013}. However, the exact value of this upper limit in
metallicity for the formation of C stars is highly sensitive to
uncalibrated model details.

Here, we provide the first observations of TP-AGB stars in a
metal-rich environment (M31) capable of providing comprehensive model
calibrations of the third dredge up and the C star metallicity ceiling
as a follow-up to the pilot program, described in
\citet{Boyer+2013}. In this first paper, we describe the survey design
(\S\ref{sec:survey}), the observations and stellar classification
(\S\ref{sec:data}), and describe the resulting C/M in the context of
the other nearby galaxies (\S\ref{sec:results} and
\S\ref{sec:disc}). We find that the carbon star formation efficiency
decreases at a rate higher than expected based on observations in more
metal-poor galaxies, but that the predicted metallicity ceiling for
carbon star formation remains elusive. A forthcoming paper (Chen et
al., in preparation) will explore the resulting constraints to the
Padova TP-AGB models.

\begin{deluxetable*}{rcrrrccccc}
  \tablewidth{0pc}
  \tabletypesize{\normalsize}
  \tablecolumns{10}
  \tablecaption{Field Information \& C and M Star Statistics\label{tab:fields}}
  \tablehead{
    \colhead{Field} &
    \colhead{PHAT Brick\tablenotemark{a}} &
    \colhead{RA} &
    \colhead{Dec} &
    \colhead{PA} &
    \colhead{$[M/H]$\tablenotemark{b}} &
    \colhead{$R_{\rm deproj}$} &
    \colhead{$N_{\rm C}$\tablenotemark{c}} &
    \colhead{$N_{\rm M}$\tablenotemark{c}} &
    \colhead{$N_{\rm spec}$\tablenotemark{d}} \\
    \colhead{} &
    \colhead{} &
    \colhead{(J2000)} &
    \colhead{(J2000)} &
    \colhead{(E of N)} &
    \colhead{} &
    \colhead{ (kpc)} &
    \colhead{} &
    \colhead{} &
    \colhead{}
    }
  \startdata
Pilot (0) & 3 & 00:43:21.66 & $+$41:21:55.1 & 225\fdg0 & $+0.058\pm0.010$& 2.61& 9& 2511& 0\\
    1 & 5 & 00:43:18.49 & $+$41:26:10.9 &  67\fdg0 &  $+0.053\pm0.011$   & 2.83&  7&1971& 0\\
    2 & 5 & 00:43:18.33 & $+$41:29:29.3 &  63\fdg5 &  $+0.027\pm0.017$   & 4.17&  5&1254& 0\\
    3 & 7 & 00:43:31.67 & $+$41:32:28.3 &   3\fdg3 &  $+0.013\pm0.019$   & 4.83& 11&1224& 0\\
    4 & 7 & 00:43:40.58 & $+$41:34:56.8 &  35\fdg2 &  $+0.000\pm0.022$   & 5.50&  6& 948& 0\\
    5 & 5 & 00:43:48.18 & $+$41:26:22.4 &  65\fdg1 &  $+0.023\pm0.017$   & 4.34&  4&1847& 0\\
    6 & 7 & 00:44:07.64 & $+$41:30:09.3 &  65\fdg0 &  $-0.001\pm0.022$   & 5.56& 10&1194& 0\\
    7 & 4 & 00:43:51.57 & $+$41:19:12.6 &  32\fdg4 &  $-0.030\pm0.028$   & 7.00& 22&1135& 0\\
    8 & 6 & 00:44:08.23 & $+$41:22:22.1 &  66\fdg8 &  $-0.045\pm0.031$   & 7.77& 20& 961& 0\\
    9 & 6 & 00:44:17.65 & $+$41:25:08.6 &  33\fdg8 &  $-0.048\pm0.032$   & 7.88& 21& 933& 2\\
   10 & 8 & 00:44:39.27 & $+$41:29:32.6 &  70\fdg0 &  $-0.069\pm0.036$   & 8.96& 31& 731& 0\\
   11 & 9 & 00:44:13.99 & $+$41:37:10.6 &  34\fdg7 &  $-0.012\pm0.024$   & 6.08&  9& 894& 1\\
   12 & 10& 00:45:02.84 & $+$41:36:00.3 &  20\fdg0 &  $-0.086\pm0.039$   & 9.82& 23& 686& 0\\
   13 & 11& 00:44:57.09 & $+$41:40:46.4 &  33\fdg1 &  $-0.063\pm0.035$   & 8.64& 12& 688& 0\\
   14 & 13& 00:44:56.13 & $+$41:46:17.2 & 272\fdg0 &  $-0.066\pm0.035$   & 8.82& 16& 578& 0\\
   15 & 15& 00:44:55.97 & $+$41:53:42.8 &  29\fdg0 &  $-0.101\pm0.042$   &10.55& 24& 578& 1\\
   16 & 15& 00:45:17.43 & $+$41:51:13.7 & 342\fdg5 &  $-0.095\pm0.041$   &10.25& 35& 627& 1\\
   17 & 14& 00:45:29.18 & $+$41:46:11.9 &  33\fdg1 &  $-0.105\pm0.043$   &10.75& 22& 590& 1\\
   18 & 17& 00:45:31.47 & $+$41:57:16.3 & 338\fdg0 &  $-0.123\pm0.047$   &11.67& 19& 448& 0\\
   19 & 16& 00:45:42.86 & $+$41:53:48.8 & 273\fdg0 &  $-0.123\pm0.047$   &11.63& 26& 454& 0\\
   20 & 18& 00:46:15.94 & $+$41:57:27.1 & 280\fdg0 &  $-0.164\pm0.055$   &13.68& 14& 189& 5\\
  \enddata

  \tablecomments{\ The pilot program ID is GO-12862, obtained in
    September 2012; the remaining 20 fields are from GO-14072,
    obtained between July 2016 and Feb 2017. The WFC3/IR field of view
    is 2\farcm1 $\times$ 2\farcm3. The DOI for this dataset is: 10.17909/t9-9xwn-et04.}

  \tablenotetext{a}{\ PHAT Brick \citep{Dalcanton+2012b}}.

  \tablenotetext{b}{\ Metallicities are derived using the gradient
    measured by \citet{Gregersen+2015}. The uncertainties reflect the
    uncertainty in the gradient slope ($\pm$0.004 dex/kpc). The
    absolute metallicities are subject to systematics and assumptions
    and may vary by about 0.3~dex consistently across all fields.}

  \tablenotetext{c}{\ $N_{\rm C}$ and $N_{\rm M}$ are the number of C
    and M-type giants, respectively (\S\ref{sec:class}).}
  
    \tablenotetext{d}{\ The number of carbon stars in the field
      identified via optical spectroscopy from \citet{Hamren+2015}.}
    
\end{deluxetable*}

\section{Survey Design}
\label{sec:survey}

We imaged 20 fields within the PHAT footprint with the WFC3/IR camera
on the Hubble Space Telescope (HST) between July 2016 and Feb 2017
(GO-14072).  Figure~\ref{fig:map} displays a map of the fields, which
we placed to sample a wide range in metallicity and age
(\S\ref{sec:survey}), while also covering locations with
spectroscopically-confirmed AGB stars \citep{Hamren+2015}. We list the
field information in Table~\ref{tab:fields}, including an estimate of
the deprojected radius which we compute assuming a position angle of
38\degr, inclination angle of 74\degr \citep{Barmby+2006}, central
position of ${\rm R.A.}=10$\fdg$68458$ and ${\rm Dec.} = 41$\fdg$2692$
\citep{McConnachie+2005}, and distance of 776 kpc
\citep{Dalcanton+2012b}.

Data from the pilot program \citep{Boyer+2013} are also included here,
and marked either as `pilot' or `Field 0' in tables and figures
(outlined in red in Figure~\ref{fig:map}). Observations of the pilot
field were taken in Sep 2012 (GO-12862). All analysis here includes a
re-reduction of the pilot field, described in Section~\ref{sec:data}.

\subsection{Metallicity}
\label{sec:Z}

To study how metallicity affects C/M across M31's disk, we adopt
metallicities from \citet{Gregersen+2015}. That work estimated [M/H]
by interpolating isochrones over 7 million red giant branch (RGB)
stars using the Padova PARSEC1.2s models \citep{Bressan+2012,
  Chen+2014, Tang+2014} and spatially binning the results. They find
an absolute offset compared to the BaSTI models
\citep{Pietrinferni+2007}, with the BaSTI models typically resulting
in a metallicity higher by 0.3~dex over the full metallicity
range.

\citet{Gregersen+2015} investigated the age-metallicity degeneracy by
testing different fiducial ages with and without age gradients in the
disk. They find that the slope of the metallicity gradient is not
affected by an absolute offset in stellar age, but it shifts slightly
(by $\approx$$-0.003$~dex/kpc) if they assume a typical stellar age
gradient ($<$$\mid$0.1$\mid$~Gyr/kpc).  The absolute metallicity is
anticorrelated with RGB age, decreasing by $\approx$0.2~dex with a
fiducial age of 12 Gyr. This, along with the offset found using the
BaSTI models, suggests that while the relative metallicities between
fields are reliable, the absolute values may vary by
up to $\pm$0.3~dex.

While the absolute metallicity is somewhat uncertain, the spatial
gradient in metallicity appears to be more robust. Overall, the
\citet{Gregersen+2015} analysis found a metallicity gradient of
$-0.020 \pm 0.004$ dex/kpc from $\sim$ 4--20 kpc, assuming a mean RGB
age of 4 Gyr. The \citet{Gregersen+2015} slope is consistent
with the slopes measured using \ion{H}{2} regions over the same region
of the disk \citep{ZuritaBresolin2012, Sanders+2012}, suggesting that
the gradient is relatively stable over a long
timescale. \citet{Sanders+2012} find no abundance gradient from
measurements of planetary nebulae (PNe), a population with a similar
age to TP-AGB stars. However, they note that selection effects exclude
higher metallicity PNe from their sample and conclude that the
inclusion of metal-rich PNe would likely result in a metallicity
gradient.

We extrapolate the metallicity gradient to the innermost fields that
were not analyzed in \citet{Gregersen+2015} due to crowding and
incompleteness ($\lesssim$3~kpc). In these inner regions,
\citet{Saglia+2018} estimated the metallicities based on spatial maps
of absorption line features fit with simple stellar population (SSP)
models. Their spatial coverage reaches the position of our pilot
program, for which they find ${\rm [M/H]} = -0.05$ to $+0.05$,
consistent with the extrapolated value from the \citet{Gregersen+2015}
gradient. We list the adopted metallicities in
Table~\ref{tab:fields}.

\subsection{Age}
\label{sec:age}

TP-AGB stars have initial masses of $\approx$0.8--8~$M_\odot$. All
TP-AGB stars are initially M-type, while a subset eventually become
carbon stars once enough carbon has dredged up to the surface. At
M31's metallicity, carbon TP-AGB stars are predicted to form only
between ages of approximately 200~Myr and 4~Gyr
\citep[$\approx$1.5--4.5~$M_\odot$;][]{Karakas+2016}, while the rest
of the TP-AGB stars remain M-type.  Therefore, we can only expect the
C/M ratio to be a useful diagnostic if the stellar population spans the
appropriate age range.

\citet{Lewis+2015} and \citet{Williams+2017} estimated the
spatially-resolved star-formation history (SFH) across the disk by
fitting stellar evolution models to PHAT color-magnitude diagrams
(CMDs). They find that there are global star formation events from
$\approx$1.5-4~Gyr and at ages $\gtrsim$8~Gyr. The star-forming rings
(Fig.~\ref{fig:map}) also contain a young component $\lesssim$900 Myr
old. All of our fields contain a population from the global
$\approx$1.5--4~Gyr event \citep{Williams+2017}, which produced carbon
stars with initial masses around 1.5--2.1~$M_\odot$.  We expect the
bulk of the carbon stars in M31 to have formed during this event.

The ancient stellar population in M31, which is smoothly distributed
from the bulge to the outer disk will contribute very few carbon
stars. Since the bulge is dominated by this ancient population, we
placed our innermost fields at a distance that minimizes the
contribution of the bulge while also reaching the highest possible
metallicity.  Bulge stars are also enriched in He
\citep[e.g.][]{Nataf+2011,Rosenfield+2012}, possibly due to pollution
from an earlier population. He enrichment decreases the likelihood of
carbon star formation even further by decreasing the stellar lifetime,
resulting in fewer thermal pulses and less carbon dredge up
\citep{Karakas2014}. The PHAT optical CMDs (see \S\ref{sec:phot})
indicate that all of our fields, including those nearest the bulge,
contain a strong contribution from a young population
\citep[also,][]{Lewis+2015}. This suggests that
the inner disk is a strong component in all of our inner fields, and
thus carbon stars are expected to be present. In addition, we noted in
\citet{Boyer+2013} that the bulge is expected to contribute $<$30\% of
the integrated $I$-band light in our pilot field
\citep{Dorman+2012}. Since the $I$-band is the least susceptible to
stochastics from bright objects \citep{MelbourneBoyer2013}, the young
inner disk is therefore the dominant population even in our
most central fields.

\subsection{Known AGB Stars}
\label{sec:size}

The Spectroscopic and Photometric Landscape of Andromeda's Stellar
Halo \citep[SPLASH;][]{Dorman+2012} program obtained thousands of
optical spectra ($\lambda \sim 5000$--$9000$~\AA) in the disk using
the DEIMOS spectrograph on the Keck~II telescope. They found a total
of 103 carbon stars and 736--1605 M stars, depending on how they are
classified, based on the presence of CN and TiO absorption features
\citep{Hamren+2015}.  Our fields contain 11
spectroscopically-confirmed carbon stars from the SPLASH survey, which
provide a useful test of our photometric classification. These stars
are marked in Figures~\ref{fig:map} and \ref{fig:mainccd} and noted in
Table~\ref{tab:fields}.

\section{Data \& Analysis}
\label{sec:data}

\subsection{Observations}
\label{sec:obs}

Each field in Table~\ref{tab:fields} was imaged with the F127M, F139M,
and F153M filters on WFC3/IR. These filters are narrow enough ($\Delta
\lambda = 64$--69~nm) to sample molecular features in TP-AGB stars and
enable classification into C- and M-type giants
\citep{Boyer+2013,Boyer+2017}.

The WFC3/IR detector reads out continuously and non-destructively
(multiaccum mode\footnote{WFC3 handbook: http://www.stsci.edu/hst/wfc3/documents/\newline handbooks/currentIHB/wfc3\_cover.html}). Here, we use the
{\asciifamily SPARS50} readout pattern, which samples the exposure
ramp linearly with 50s intervals. Each field is dithered to improve
pixel sampling and mitigate effects from artifacts and/or cosmic
rays. The exposures follow either a 2-pt or 4-pt {\asciifamily
  WFC3-IR-DITHER-BOX-MIN} pattern, which provides optimal PSF
sampling. The total combined exposure time is 861.7s for F127M, and
855.9~s each for F139M and F153M.

\subsection{Photometry \& Crowding}
\label{sec:phot}

We used the PHAT photometry pipeline, described in
\citet{Williams+2014}. Briefly, we apply the point spread function
(PSF) fitting package DOLPHOT \citep{Dolphin2000} to each of the
21 near-IR fields.  First, each CCD readout of each exposure in all of the
overlapping PHAT ACS/WFC and WFC3/IR data were run through the DOLPHOT
PSF-fitting photometry routine along with each CCD readout of the new
data.  These single exposure catalogs were then put through an
alignment routine to find a common astrometric solution for all of the
exposures.  This solution was applied to all of the headers in the
original flat-fielded and CTE-corrected frames ({\tt flc} files for
WFC exposures and {\tt flt} files for IR exposures).  The exposures
with updated astrometry were then run through the {\tt PyRAF} routine
{\tt astrodrizzle} to generate deep stacked images and updated pixel
masks.  The pixels masks and area corrections were then applied to
each CCD readout.  Finally, these individual CCD readouts were
simultaneously measured using the DOLPHOT PSF fitting routine, which
finds stars using the full stack of all overlapping pixels in memory
and forcing fits of the PSF at that sky location in every CCD read.
All of these measurements were then combined in each band to optimize
photometric depth.

The resulting photometry catalogs contain the measured brightness of
each detected star along with several quality metrics of the
measurement.  These quality metrics were then used to clean the
catalog of unreliable measurements using the {\tt gst} criteria
described in Williams et al. (2014).  However, we found that the F127M
filter contained some spurious detections with large {\tt chi} values
in the DOLPHOT output.  Thus, we applied an additional criterion of
{\tt chi} $<4$ for the F127M band.

\begin{figure}
  \includegraphics[width=\columnwidth]{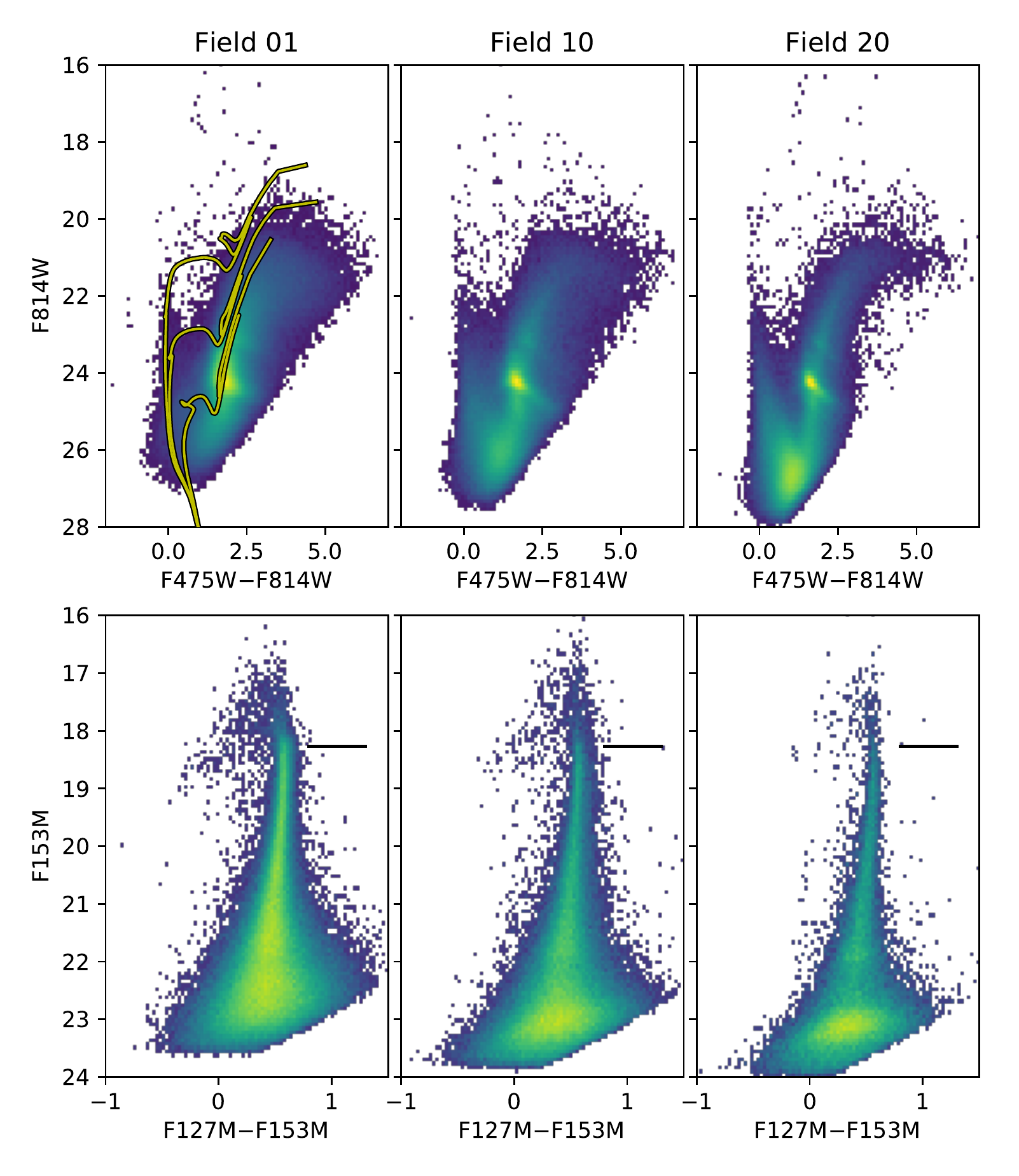}
  \caption{Sample optical and IR medium-band color-magnitude diagrams
    for the innermost field, outermost field, and one intermediate
    field. The optical CMDs (upper panels) illustrate both the
    variation in age and differential extinction. Parsec v1.2S
    isochrones \citep{Bressan+2012} with $\log(t_{\rm age}/{\rm yr}) = $8, 8.5,
    and 9 and ${\rm [M/H]} = 0.119$ are included in the left panel
    (yellow lines). Differential extinction extends the RGB width
    towards the lower right. The IR CMDs (lower panels) illustrate a
    weaker impact from differential extinction. TP-AGB stars are
    mostly located above the tip of the RGB (TRGB), marked by a horizontal black line
    (\S\ref{sec:trgb}), and extend towards fainter magnitudes at blue
    colors.}
  \label{fig:cmds}
\end{figure}

Foreground extinction is low towards M31, especially in the
near-IR. Nevertheless, all photometry presented in this paper is
corrected for a mean foreground extinction using $A_{\rm V}=0.1922$,
derived from the \citet{Schlegel+1998} map and assuming $R_{\rm V} =
3.1$. We adopt $A_\lambda$ values listed in Table~\ref{tab:av},
computed as in \citet{Girardi+2008} using the \citet{Cardelli+1989}
extinction curve with $R_{\rm V} = 3.1$. We also verify that these
coefficients change little (less than 2\%) for stars covering a 2500~K range
in $T_{\rm eff}$.

\begin{deluxetable}{rc}
  \tabletypesize{\normalsize}
  \tablecolumns{2}
  \tablecaption{Adopted $A_{\lambda}$\label{tab:av}}
  \tablehead{
    \colhead{\hspace{0.1in}Filter} & 
    \colhead{\hspace{0.1in}$A_{\lambda}/A_{V}$} \\
    \colhead{}&
    \colhead{(mag)}
    }
  \startdata
  F475W & 1.185\\
  F814W & 0.610\\
  F127M & 0.274\\
  F139M & 0.240\\
  F153M & 0.204\\
  F110W & 0.337\\
  F160W & 0.204\\
  \enddata
\end{deluxetable}

Example optical and IR medium-band CMDs are shown in
Fig.~\ref{fig:cmds}, including the innermost field, outermost field,
and one intermediate field. The optical CMDs illustrate the relative
ages of the fields, with $\log (t_{\rm age}/{\rm yr})$ = 8, 8.5, and 9 Parsec
v1.2S isochrones included in the left panel \citep{Bressan+2012}. The
optical CMDs also illustrate the differential extinction, with the RGB
width increasing towards the lower right, most visible in the red
clump at F814W$\approx$24~mag; the extinction is lowest in the outer
fields (e.g., Field 20 in the right panels).  The medium-band CMDs are
similar in all fields, with a smaller impact from differential
extinction.  TP-AGB stars occupy magnitudes brighter than
F153M$\approx$18~mag and extend to fainter magnitudes at blue colors
(\S\ref{sec:trgb} \& \ref{sec:class}).

\begin{figure}
  \includegraphics[width=\columnwidth]{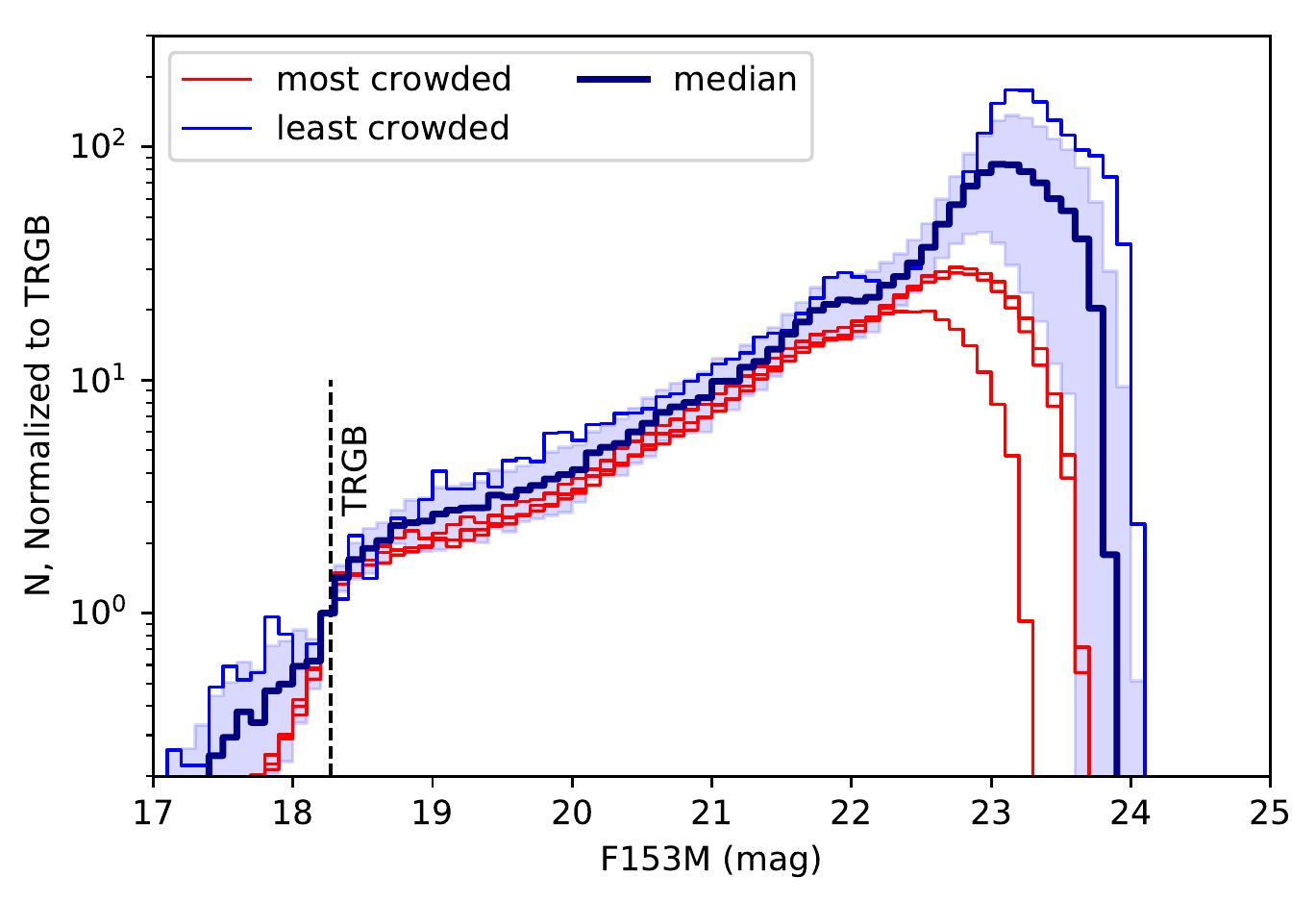}
  \caption{Median F153M luminosity function and 1-$\sigma$ range
    (shaded region), normalized to the TRGB.  Only 3 fields (the
    pilot, and fields 01 and 05; in red) have luminosity functions
    that fall below the shaded region, indicating a higher level of
    incompleteness at faint magnitudes due to crowding. The thin blue
    line shows field 20, which is at the largest galactic radius and
    should therefore have the least amount of crowding among our
    fields. }
  \label{fig:crowd}
\end{figure}

Our most crowded field is the pilot field at 2.6~kpc, where the
stellar density for stars with $18.5 < {\rm F160W} < 19.5$~mag ranges
from 1.8--3 per arcsec$^2$. In this regime, crowding dominates over
other photometric uncertainties. \citet{Williams+2014} showed that,
for this stellar density in the PHAT data, the photometry begins to be
biased slightly towards brighter magnitudes near 19--20~mag in F160W
and the RMS uncertainty is near 0.2~mag, with slightly smaller values
in F110W. We expect crowding to affect the photometry at similar
limits in our most crowded fields, with a much weaker effect in
regions farther out in the disk.  Figure~\ref{fig:crowd} shows the
median and 1-$\sigma$ luminosity functions for F153M, the filter with
the lowest resolution and strongest crowding. The
crowding level is similar in most of our fields, with photometric
completeness beginning to decrease around F153M = 22--23~mag, several
magnitudes below the tip of the red giant branch (TRGB;
\S\ref{sec:trgb}).  Only 3 fields (the pilot, 01, and 05) are more
crowded than the 1-$\sigma$ luminosity function (red lines in
Fig.~\ref{fig:crowd}), but even these are complete to $\gtrsim$3~mag
fainter than the TRGB. Based on \citet{Williams+2014}, we expect the
RMS error to be 0.1~mag or less in all but the inner $\sim$3 fields
(in crowding order: the pilot field, field 1, and field 5). Since our
C and M star selection criteria (\S\ref{sec:class}) are restricted to stars
brighter than the TRGB and well above the completeness limit, the
uncertainties on the number of AGB stars are dominated by Poisson
statistics rather than crowding or sensitivity limits.

\subsection{TRGB}
\label{sec:trgb}

We must first select the TP-AGB stars before we can identify C- and
M-giant stars. We do this by selecting stars that are brighter than the
TRGB in at least one of the near-IR filters.

The RGB in our data is well-populated and unaffected by incomplete
photometry to approximately 3~mag below the TRGB. We therefore follow
the procedure described in \citet{Mendez+2002} to measure the TRGB in
each filter. We begin by constructing a luminosity function using a
Bayesian optimization method to select the bin
size.\footnote{http://docs.astropy.org/en/stable/api/\\astropy.stats.knuth\_bin\_width.html}
To minimize the contributions from foreground and main sequence stars
to the luminosity function, we only include stars with ${\rm
  F127M}-{\rm F153M} > 0.1$~mag. Next, we pass the Gaussian-smoothed
luminosity function through a Sobel edge-detection filter to detect
the TRGB, i.e., where the number of sources drops. We perform 500
Monte Carlo resampling trials to test the effect of random photometric
errors and random variations in the bin sampling of the luminosity
function. The resulting TRGBs, measured using all 21 fields
simultaneously to avoid stochastics, are listed in
Table~\ref{tab:trgb}.

The near-IR TRGB is known to vary with age and metallicity
\citep[e.g.,][]{Dalcanton+2012a,McQuinn+2017,McQuinn+2019}, and thus
may not be constant across M31's disk. In Table~\ref{tab:trgb}, we
include the difference in the TRGB ($\Delta$TRGB) measured from the two
innermost fields ($\sim$2.7~kpc, ${\rm [M/H]}\sim+0.05$) and two outer
fields ($\sim$11.6~kpc, ${\rm [M/H]}\sim-0.12$). Given the relatively
small range in metallicity, the TRGB difference between these two
extremes is likewise small in all filters, less than 0.05~mag and
typically on the order of the 1-$\sigma$ uncertainties. By using the
TRGB in all five filters to select the initial sample of stars, we
avoid any biases caused by differences in environment.

Most TP-AGB stars are brighter than the TRGB, though some can be
fainter for the following reasons: (1) they are at the luminosity dip
that characterizes the initial fraction of a thermal pulse cycle
\citep{BoothroydSackmann1988}; (2) they have enough circumstellar dust
to cause self-extinction; and/or (3) they have strong molecular
absorption, with the particular molecule dependent on the filter. In
our case, water absorption in M-giant stars is the dominant effect
causing TP-AGB stars to be fainter than the TRGB, since the broad
water feature at 1.4~\micron\ is covered to some degree by F139M,
F153M, F110W, {\it and} F160W \citep[see Fig.~1
  in][]{Boyer+2013}. F127M has the least overlap with this water
feature and inspection of our data confirms that it has the fewest
TP-AGB stars below its TRGB.

\begin{deluxetable}{rcccc}
  \tabletypesize{\normalsize}
  \tablecolumns{5}
  \tablecaption{TRGB\label{tab:trgb}}
  \tablehead{
    \colhead{Filter} &
    \colhead{$\lambda_{\rm pivot}$} &
    \colhead{TRGB} &
    \colhead{1\,$\sigma$}&
    \colhead{$\Delta$TRGB}\\
    \colhead{}&
    \colhead{$\mu$m} &
    \colhead{(mag)}&
    \colhead{(mag)}&
    \colhead{(mag)}
    }
  \startdata
  F127M & 1.27 & 18.80 & 0.04 & 0.05\\
  F139M & 1.38 & 18.66 & 0.03 & 0.03\\
  F153M & 1.53 & 18.27 & 0.01 & 0.05\\
  F110W & 1.15 & 19.28 & 0.02 & 0.01\\
  F160W & 1.54 & 18.28 & 0.02 & 0.01\\
  \enddata

  \tablecomments{\ The TRGB is measured using all 21 fields
    simultaneously. The $\Delta$TRGB measurement is the absolute
    difference in the TRGB measured in two outer fields combined
    (Fields 18 \& 19) and the two innermost fields combined (Fields 0
    \& 1); see text. Pivot wavelength is the effective filter wavelength, as
    quoted in the WFC3/IR handbook.}
 
\end{deluxetable}

\subsection{C and M Star Classification}
\label{sec:class}

\begin{figure}
  \includegraphics[width=\columnwidth]{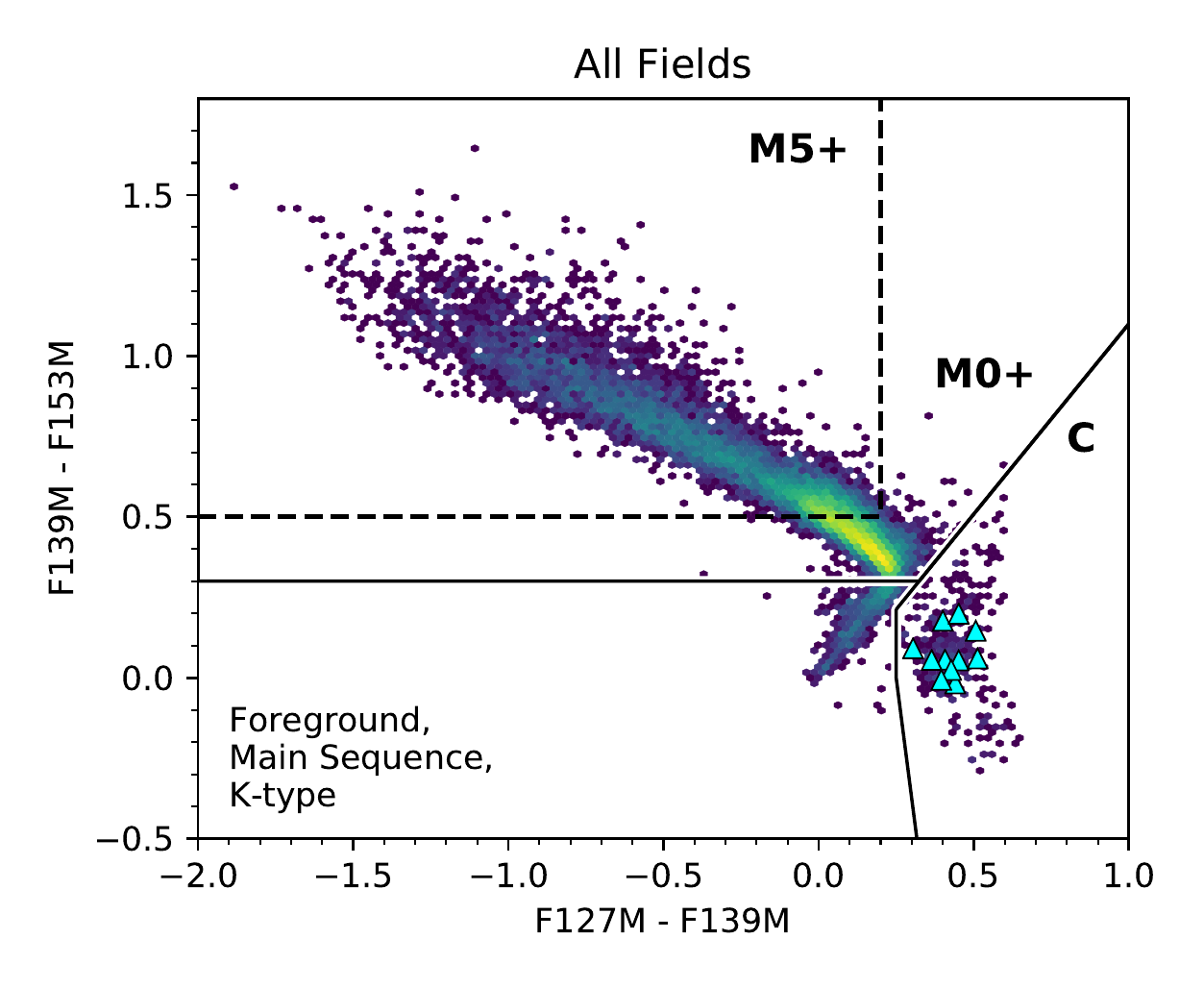}
  \caption{Medium-band color-color diagram showing the separation of C
    and M giant stars (solid line). We include only the 21,460 stars
    that are brighter than the TRGB in at least one of the filters
    listed in Table~\ref{tab:trgb}. Objects in the lower left box
    include contaminating objects (foreground, main sequence stars,
    and K-type AGB stars). Water absorption in M stars increases along
    the sequence towards the upper left, with late-type M giants
    showing the strongest water absorption (stars of type later than
    M5 are marked with a dashed line). Atmospheric C/O in C stars
    approximately increases towards the lower right. Circumstellar
    dust moves stars up and to the right while maintaining the
    separation between C and M stars \citep{Boyer+2017}. Known C stars
    from the SPLASH program are marked by cyan triangles
    \citep{Hamren+2015}.}
  \label{fig:mainccd}
\end{figure}

To identify C- and M-type giants, we start by isolating the bulk of
the TP-AGB stars by selecting stars that are brighter than the TRGB in
at least one of the filters listed in Table~\ref{tab:trgb}. Next, we
use the color-color diagram (CCD) in Figure~\ref{fig:mainccd} to
divide the TP-AGB stars into C-type and M-type. This technique is
described in detail in \citet{Boyer+2013,Boyer+2017}; it takes
advantage of water absorption in M-giant stars and CN$+$C$_2$
absorption in carbon stars to create a remarkably clean separation of
these stellar types in the medium-band filters. We largely use the
color cuts from \citet{Boyer+2017}, illustrated in
Figure~\ref{fig:mainccd}, which isolate all M stars later than type
M0. Main sequence, foreground, and K-type stars in our sample are
removed by excluding objects on the lower-left region of the
CCD. M31's high metallicity leads its stars to have lower stellar
temperatures, favoring the formation of cooler M-type stars over
K-type stars on the AGB. As a result, many of the M31 CCDs show a
clear gap between M-type and K-type stars near F139M$-$F153M = 0.3~mag that was not seen in the
metal-poor galaxies from \citet{Boyer+2017} (e.g., Fields 0, 1, 2, 4, 6,
8, 14, \& 20 in Figs.~\ref{fig:appendix_ccd1} and
\ref{fig:appendix_ccd2}). We use this gap to slightly adjust the M
star box down from the ${\rm F139M-F153M} = 0.35$~mag adopted by
\citet{Boyer+2017} to ${\rm F139M-F153M} =
0.3$~mag. Table~\ref{tab:fields} shows the resulting C and M star
counts for each field.

The TRGB cuts used to select the initial subset of stars in
Figure~\ref{fig:mainccd} may be missing a small fraction of TP-AGB
stars. For example, the dustiest objects are faint or even undetected
at 1.5~\micron\ from self extinction.  However, these objects are
$\lesssim$5\% of the total TP-AGB population and thus have only a
small effect on C/M \citep{Blum+2006,Boyer+2011,Boyer+2017}. The
fraction of very dusty stars does not appear to change much with
metallicity, with both the LMC and SMC showing similar fractions, and
thus M31 is not expected to be strongly affected. In addition, models
suggest that dust production by C stars may be independent of
metallicity \citep[e.g.,][]{Nanni+2013}, also removing concerns that
M31's stars would be preferably dust enshrouded. Ideally, Mid-IR data
would produce a complete TP-AGB census that includes all
dust-producing stars, but in practice most estimates of C/M throughout
the Local Group \citep[e.g,.][and references
  therein]{Brewer+1995,BattinelliDemers2005,Boyer+2017,Cioni2009} use
optical and/or near-IR data, and therefore also exclude the most
extreme dusty objects that could be missing from this study of M31.

\begin{figure}
  \includegraphics[width=\columnwidth]{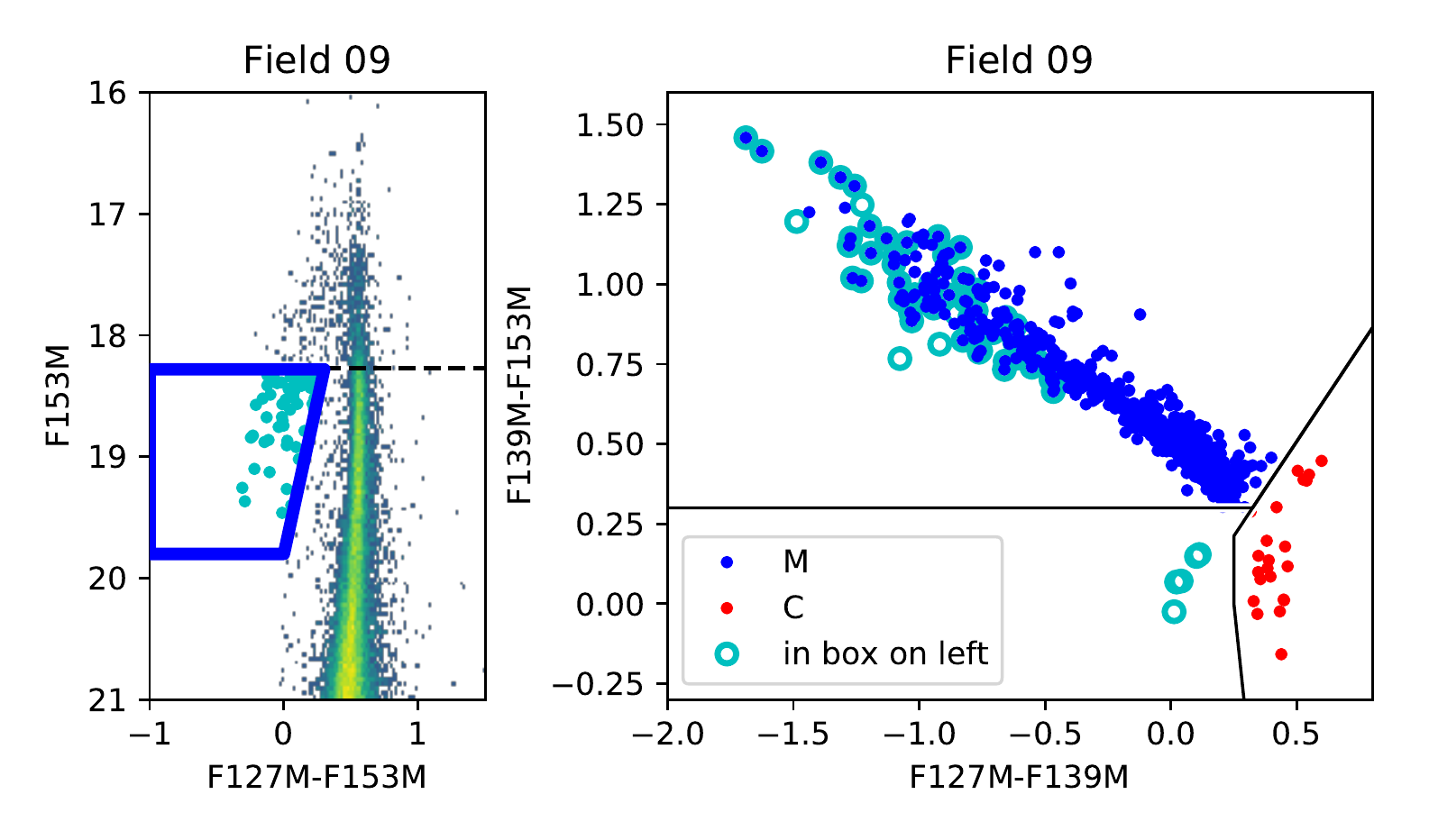}
  \caption{{\it Left:} We recover objects that are fainter than all
    five TRGBs due to deep water absorption by checking objects in the
    blue box, shown here for Field 09. The dashed line is the F153M
    TRGB. {\it Right:} Objects in the blue box on the left are plotted
    as open cyan circles, along with other C and M type stars. A
    handful of M stars in each field were recovered this way.}
  \label{fig:faintbox}
\end{figure}

In addition to the dustiest objects, the starting TRGB criterion also
misses a small number of sub-TRGB M giant stars with strong water
absorption (\S\ref{sec:trgb}). To recover these stars, we investigate
sources in the blue box in Fig.~\ref{fig:faintbox}. The lower and
right-hand edges of the box are set to minimize the contamination
from main sequence, foreground, and K-type stars; remaining interlopers are eliminated
when plotted on the medium-band CCD.  In total we recover 96 sub-TRGB
M stars, which is $\lesssim$1\% of the M giant population.

\subsubsection{M Star Definition}
\label{sec:mstars}

\defcitealias{Brewer+1995}{B95}	
\defcitealias{BattinelliDemers2005}{BD05}
\defcitealias{Hamren+2015}{H15}

Different M star definitions can make it challenging to compare the
results from other C/M studies. In the metal-rich regime, different
definitions can change the measured C/M by factors of
$\sim$1.5--4. \citet{Hamren+2015} provide a comparison of the M-giant
star definitions in M31 from \citet{Boyer+2013}, \citet{Brewer+1995}, and
\citet{BattinelliDemers2005}. The \citetalias{Brewer+1995} and
\citetalias{BattinelliDemers2005} studies use a bolometric magnitude
derived from $V$, $R$, and $I$ photometry to select the M-giant stars,
as opposed to the near-IR TRGB method used here. The result is a
dramatically different sample of M stars and a wide range in C/M. The
definitions in \citetalias{Brewer+1995} and
\citetalias{BattinelliDemers2005} are defined as:

\begin{equation}
  \left.\begin{aligned}
        V-I>&1.8,\\
        I>&18.5,\\
        M_{\rm bol}<&-3.5, {\rm where}\\
        M_{\rm bol}=&I+0.3+0.34(V-I)\\
        &-0.14(V-I)^2 - (m-M)_0,
  \end{aligned}
  \right\}
  \text{\citetalias{Brewer+1995}}
\end{equation}

\begin{equation}
  \left.\begin{aligned}
        R-I>&0.9,\\
        M_{\rm bol}<&-3.5, {\rm where}\\
        M_{\rm bol}=&I+1.7-2.3(R-I)+\\
        &1.7(R-I)^2-0.4(R-I)^3-\\
        &(m-M)_0.
  \end{aligned}
  \right\}
  \text{\citetalias{BattinelliDemers2005}}
\end{equation}

\begin{figure}
  \includegraphics[width=\columnwidth]{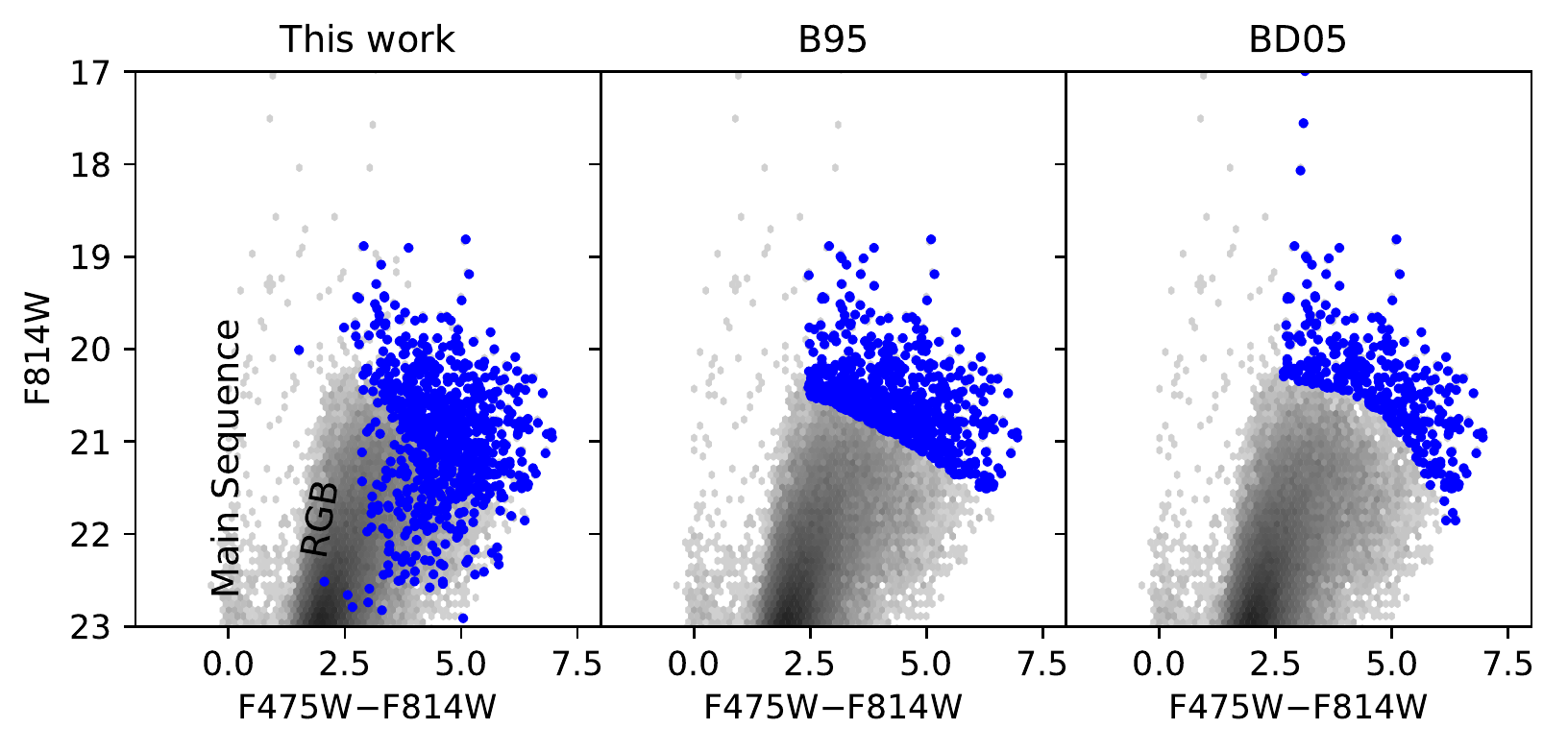}
  \includegraphics[width=\columnwidth]{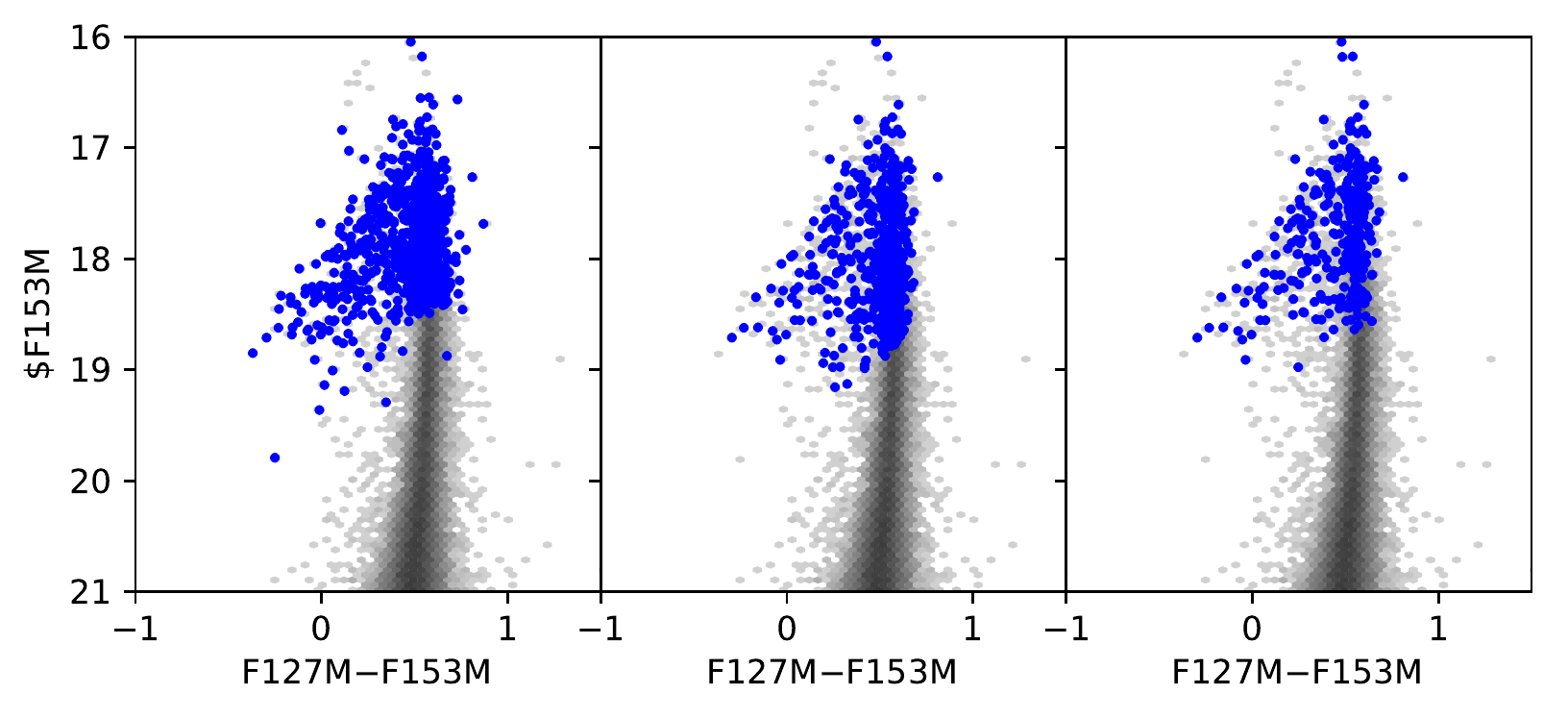}
  \includegraphics[width=\columnwidth]{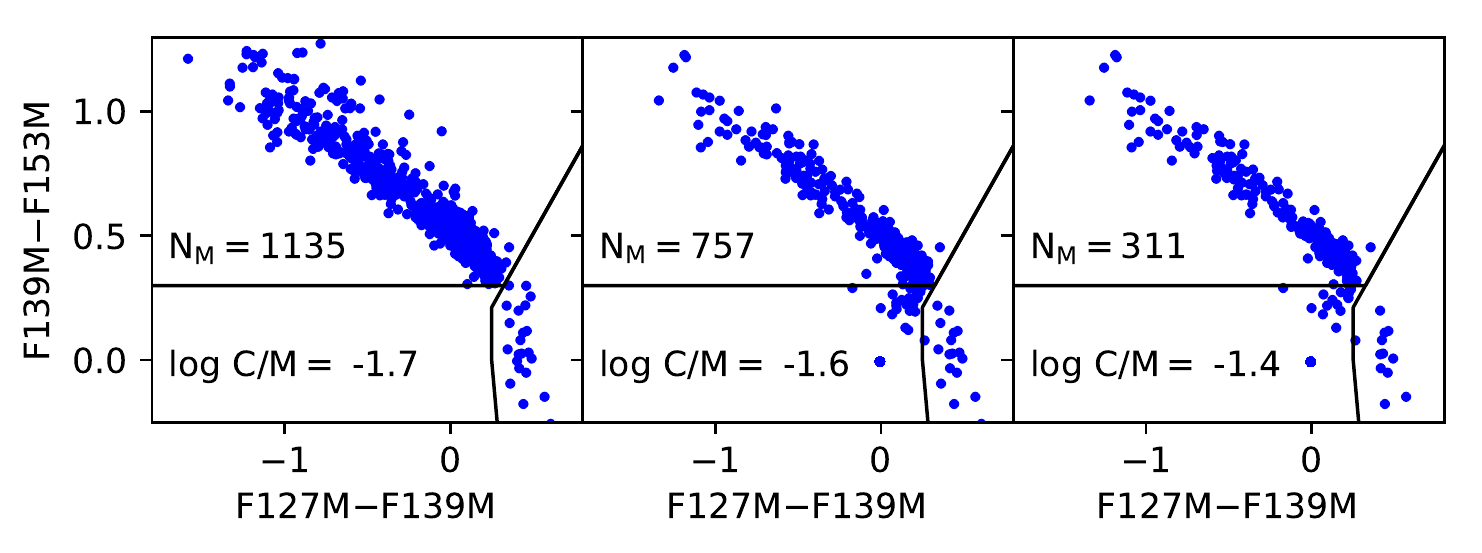}
  \caption{{\it Top:} Optical CMDs of Field 07, showing the M-giant
    population that results from the near-IR TRGB selection used here
    (left) and the $M_{\rm bol}$ selection used by
    \citetalias{Brewer+1995} and \citetalias{BattinelliDemers2005}
    (middle, right).  {\it Middle:} Same, but for the near-IR
    CMDs. {\it Bottom:} The associated medium-band HST CCD showing the
    total number of M-giant stars ($N_{\rm M}$) for each strategy. All
    blue points in each CCD would be included in a C/M ratio, with
    \citetalias{Brewer+1995} and \citetalias{BattinelliDemers2005}
    including contaminating sources in the lower left box among their M
    star numbers. By restricting to $M_{\rm bol}$ derived from optical
    filters, late-type M stars with strong optical TiO absorption (and
    thus strong near-IR H$_2$O absorption) are excluded. This
    exclusion has an effect on C/M that is worse for higher
    metallicity fields that have a higher fraction of late-type M
    giants. }
  \label{fig:mdef}
\end{figure}

The resulting M star populations are illustrated for Field 7 in
Fig.~\ref{fig:mdef}, using transformations from F814W and F475W HST
magnitudes to $V$-, $R$-,and $I$-band magnitudes \citep{Fluks+1994,
  Sirianni+2005, Hamren+2015}. By restricting the bolometric magnitude
of the sample (causing the sharp cut in blue points in the top
  middle/right panels of Fig.~\ref{fig:mdef}), a large number of
late-type M stars with strong TiO absorption at 0.4--1~\micron\ are
excluded from the cuts, resulting in a $\sim$30--60\% reduction in the
total number of M stars compared to this work. By using the near-IR
TRGB, especially in filters that are less susceptible to H$_2$O
absorption, our selection recovers these late-type M stars. The
\citetalias{Brewer+1995} and \citetalias{BattinelliDemers2005}
techniques also result in higher contamination from K-type stars, as
indicated by the higher number of sources in the lower-left boxes of
the CCDs in Figure~\ref{fig:mdef}.

Unlike in M31, late M-giant stars are rare in metal-poor environments
\citep[${\rm [M/H]} < -1$;][]{Boyer+2017}. As a result, all three M
star definitions give similar C/M values in metal-poor galaxies,
allowing for direct comparisons among multiple studies (\S\ref{sec:cm}).

\section{Results}
\label{sec:results}

Here we present our results, first the discovery of a potential new
metallicity diagnostic (C/O), then C/M across M31, which we expand on in
Section~\ref{sec:disc}.

\subsection{C/O and Metallicity}

The ratio of atmospheric carbon to oxygen (C/O) increases with each
dredge up event; if C/O$>$1, a star is considered a carbon star. While
a population at a given metallicity can span a wide range in C/O, a
metal-rich population should show a lower C/O on average due to less
efficient dredge up and a higher oxygen abundance.

\begin{figure}
  \includegraphics[width=\columnwidth]{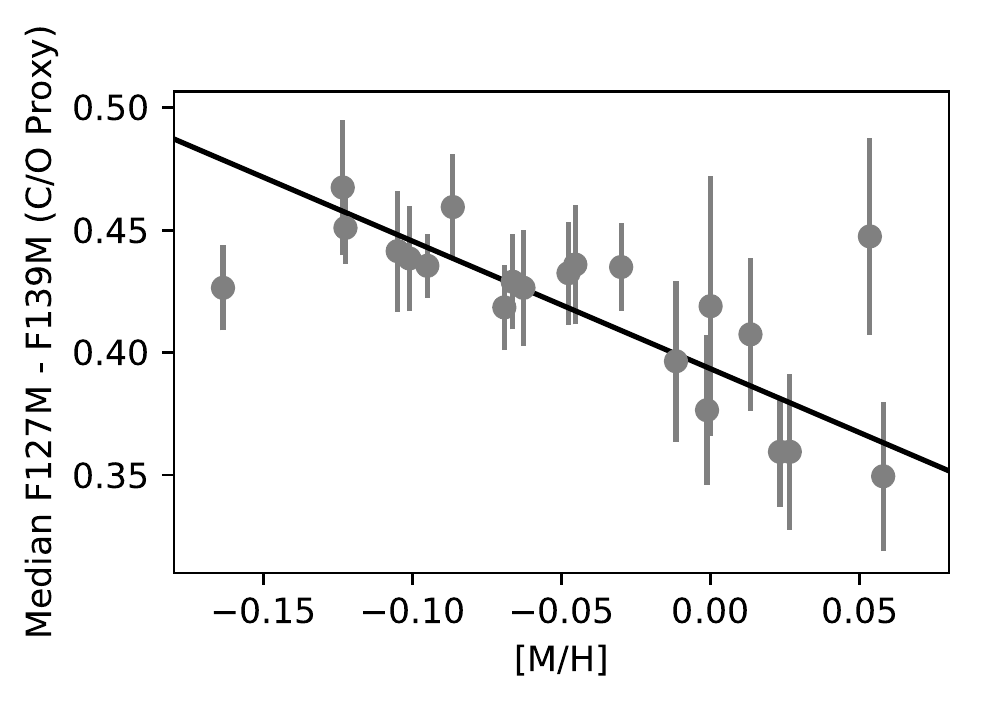}
  \caption{Atmospheric C/O proxy using the median F127M$-$F139M color
    of C stars for each field. The outliers at the high and low
    metallicity ends are Fields 1 and 20, respectively. The fitted
    line (Equation~\ref{eq:corat}) excludes those fields. Metallicity
    is from \citet{Gregersen+2015}, see Section~\ref{sec:Z}.}
  \label{fig:corat}
\end{figure}

Our medium-band filter set samples the strength of carbon features,
and thus we can potentially use the medium-band colors as an
approximate proxy for C/O in carbon stars. Carbon star atmosphere
models from \citet{Aringer+2009,Aringer+2016} predict that C/O
increases roughly for C stars from the upper left to the lower right
in the C star box of the medium-band CCD in Figure~\ref{fig:mainccd}.
We test this in figure~\ref{fig:corat}, which shows the median
F127M$-$F139M color of C stars plotted against metallicity. While the
numbers are low in the most metal-rich regions, there is a clear trend
for lower metallicity regions to have redder median colors. This
suggests that the median F127M$-$F139M color of a C star population
could be an efficient diagnostic for identifying metal-rich
populations and for tracking metallicity gradients across
galaxies. The best fit to the trend in Figure~\ref{fig:corat} is:

\begin{equation}
  \label{eq:corat}
(F127M-F139M)_{\rm med} = 0.39 - (0.52 \times {\rm [M/H]}).
\end{equation}

\noindent We note that, while both the C/O proxy and the metallicity
in Figure~\ref{fig:corat} are derived from stellar colors, these
measurements are independent. Metallicity is measured using the
optical colors of RGB stars (\S\ref{sec:Z}), and C/O is estimated
using the near-IR colors of TP-AGB stars.

This relationship may be less reliable in populations that have a high
fraction of very dusty C stars. A large dust mass veils molecular
features, causing C stars to move to the upper right of the C star box
marked in Figure~\ref{fig:mainccd}. However, most galaxies studied
show that stars this dusty comprise $\lesssim$5\% of the C star
population \citep{Boyer+2011,Srinivasan+2016,Boyer+2017}.

\subsection{C/M Across M31}
\label{sec:m31cm}

\begin{figure}
  \includegraphics[width=\columnwidth]{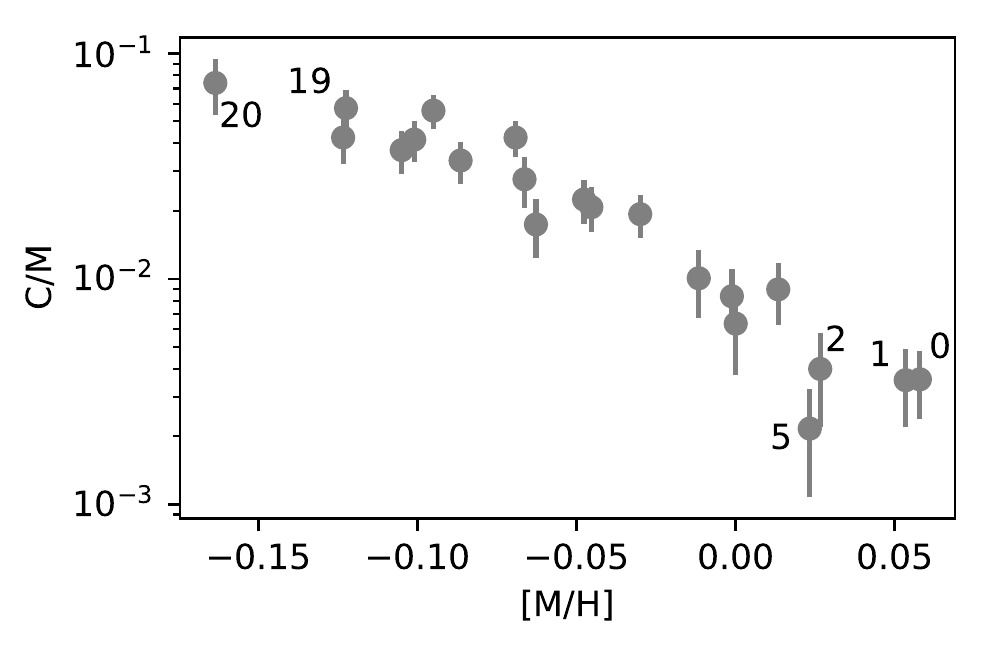}
  \caption{C/M vs. metallicity. The error bars on C/M
    reflect Poisson uncertainties. Error bars for [M/H] are not
    included for clarity, but are discussed in more detail in
    \S\ref{sec:Z}. C/M decreases smoothly as metallicity increases,
    with no hint at a sharp C star metallicity ceiling in these
    fields.  The lowest and highest metallicity fields are marked with the field number.}
  \label{fig:trend_z}
\end{figure}

TP-AGB models predict that C/M decreases with increasing metallicity
\citep{Karakas+2002,Marigo+2013,Choi+2016}. This trend is caused both
by a higher abundance of free oxygen in metal-rich stellar atmospheres
available to bind any carbon into the CO molecule and by lower dredge
up depth resulting in less carbon transported to the
surface. Figure~\ref{fig:trend_z} shows that in M31 C/M indeed
decreases smoothly with increasing metallicity in our fields. Several
other studies have also observed C/M in M31, both in the northeast
disk \citep[5--18~kpc;][]{Hamren+2015} and along the southwest axis
\citep[5--40~kpc;][]{Brewer+1995, Battinelli+2003,Nowotny+2001}. These
studies also show a decrease with C/M at lower galactocentric
distances, and therefore at higher metallicities, but with a shallower
slope than what we see in our fields. Here, we find a slope of
$\frac{\Delta \log(C/M)}{\Delta [M/H]} \sim -6.5$. When fit on their
own, the \citet{Battinelli+2003}, \citet{Brewer+1995},
\citet{Hamren+2015} data show slopes of $-2.87$, $-2.23$, and $-2.63$,
respectively.  This slope discrepancy is discussed further in
Sections~\ref{sec:cm} and \ref{sec:check}. 

\begin{figure*}
  \centering
\includegraphics[width=0.9\textwidth]{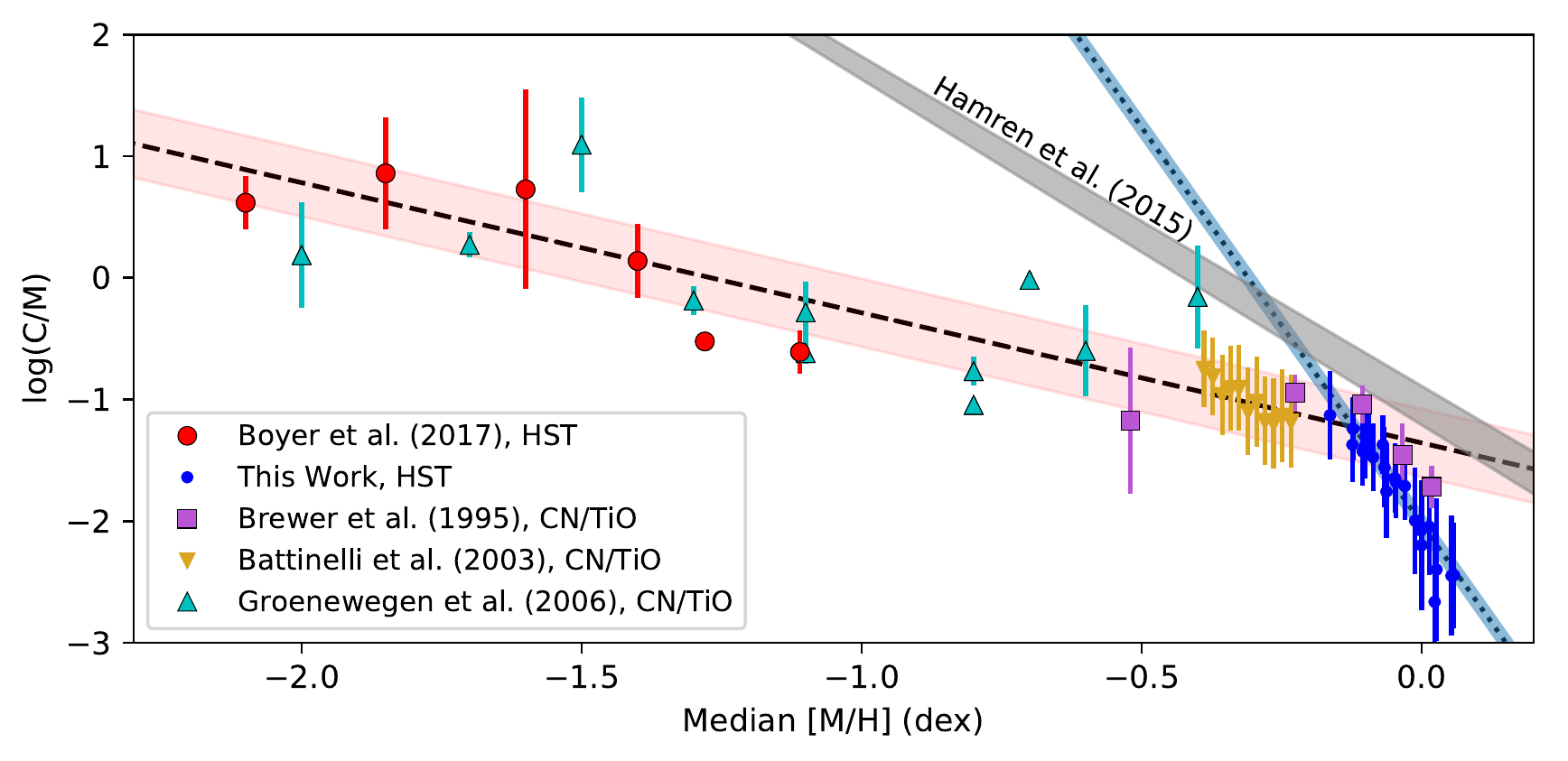}
  \caption{C/M ratio across the Local Group. Nearby star-forming dwarf
    galaxies are marked by red circles \citep{Boyer+2017} and cyan
    triangles \citep{Groenewegen2006}. M31 data on the southwest side
    of the disk are marked by purple squares \citep{Brewer+1995} and
    yellow triangles \citep{Battinelli+2003}, and the C/M from this work
    in the northeast side of the disk is marked by small blue
    dots. The red circles and blue dots were both measured using the
    HST medium-band technique to identify carbon stars, while the
    yellow triangles and cyan squares use optical narrow-band
    filters. Error bars reflect Poisson uncertainties. The dashed line
    and red shaded region mark the best fit and 1-$\sigma$ error for
    the dwarf galaxies, and the dotted line and blue shaded region
    mark the same for our fields in M31 (blue points only). The gray
    shaded region marks the C/M vs. metallicity slope measured in the
    outer disk of M31 by \citet{Hamren+2015} using spectroscopy to
    identify carbon stars. It is clear that M31 does not follow the
    trend set by the more metal-poor galaxies, indicating that the
    carbon star formation efficiency drops substantially at high
    metallicity.}
  \label{fig:lgcmr}
\end{figure*}

Most TP-AGB models also predict a metallicity ceiling above which
carbon stars cannot form \citep{Karakas+2002,Marigo+2013,Choi+2016},
though the exact metallicity of this limit has yet to be
observationally confirmed. While C/M is smoothly declining towards the
highest metallicity fields in our survey, we see no hint of a sharp
cut-off in C/M in Figure~\ref{fig:trend_z} despite the higher
metallicities reached by our fields compared to previous studies
($\sim$4--8$\times$ higher [M/H]). Nevertheless, the C/M ratio in the
innermost fields reveal an incredibly low carbon star formation
efficiency, with fewer than 10 carbon stars forming among 1000--2500 M
giant stars in Fields 0--5.

\section{Discussion}
\label{sec:disc}

\subsection{Approaching a Metallicity Ceiling?}
\label{sec:cm}

While no abrupt metallicity ceiling is apparent in
Figure~\ref{fig:trend_z}, placing the data in the context of other
nearby star-forming galaxies shows that M31 is operating in a regime
that is distinct from that of its more metal-poor counterparts. In
Figure~\ref{fig:lgcmr}, we show C/M vs. [M/H] for star-forming
galaxies from \citet{Groenewegen2006} and \citet{Boyer+2017}. The M31
data fall well below the C/M extrapolated from the more metal-poor
galaxies, by more than a factor of 10 in our innermost fields. A fit
to C/M vs. [M/H] reveals a slope of $\frac{\Delta \log({\rm
    C/M})}{\Delta {\rm [M/H]}} \sim -1.1$ for the metal-poor galaxies,
compared to a slope of $-6.5$ for the M31 fields.

This change in slope points to a shift in the dredge-up efficiency in
M31 TP-AGB stars. One result of less efficient dredge up is lower
yields of the products from TP-AGB nucleosynthesis. TP-AGB stars are
major contributors of light elements (C, Ne, N, Na, O, Mg, F, Al) and
the main contributors of $s$-process elements (Rb, Ba, Sr, La, Y, Ce,
Zr, Pb) \citep{Cristallo+2011, Fishlock+2014}. Lower yields of these
elements will affect the subsequent chemical evolution of the host
galaxy. Moreover, the severe lack of carbon stars will result in a
lack of carbon dust formation, affecting the galactic dust budget.
Having fewer C stars also has consequences for a galaxy's integrated
light, especially in the rest-frame near- and mid-IR
\citep[e.g.][]{Boyer+2011}. Given that TP-AGB stars are responsible
for up to 70\% of a galaxy's integrated light \citep{Melbourne+2012,
  MelbourneBoyer2013}, this shift in the color can have a major impact
on the appearance of galaxies.

\subsubsection{Disentangling the roles of Age and Metallicity}
\label{sec:ageZ}

The fraction of carbon stars is sensitive to the age of the population
as well as the metallicity, making the driver of the C/M trend seen in
\S\ref{sec:m31cm} uncertain. Most galaxy disks are thought to form
inside-out, resulting in both older ages and higher metallicities in
their centers. As a result, it can be challenging to decouple these
two parameters when tracking C/M across a disk. Carbon stars are
predicted to form only in a narrow age range of approximately 200~Myr
to 4~Gyr, whereas M-type TP-AGB stars can have ages anywhere from
55~Myr to 13 Gyr \citep{Karakas+2016}. These mass differences make the
C/M ratio sensitive to the relative star-formation rates over these
epochs. It is therefore possible that at least some of the trend in
C/M in Figure~\ref{fig:trend_z} could be due to a variation in age
rather than a variation in metallicity.

\begin{figure}
  \begin{center}
    \includegraphics[width=0.8\columnwidth]{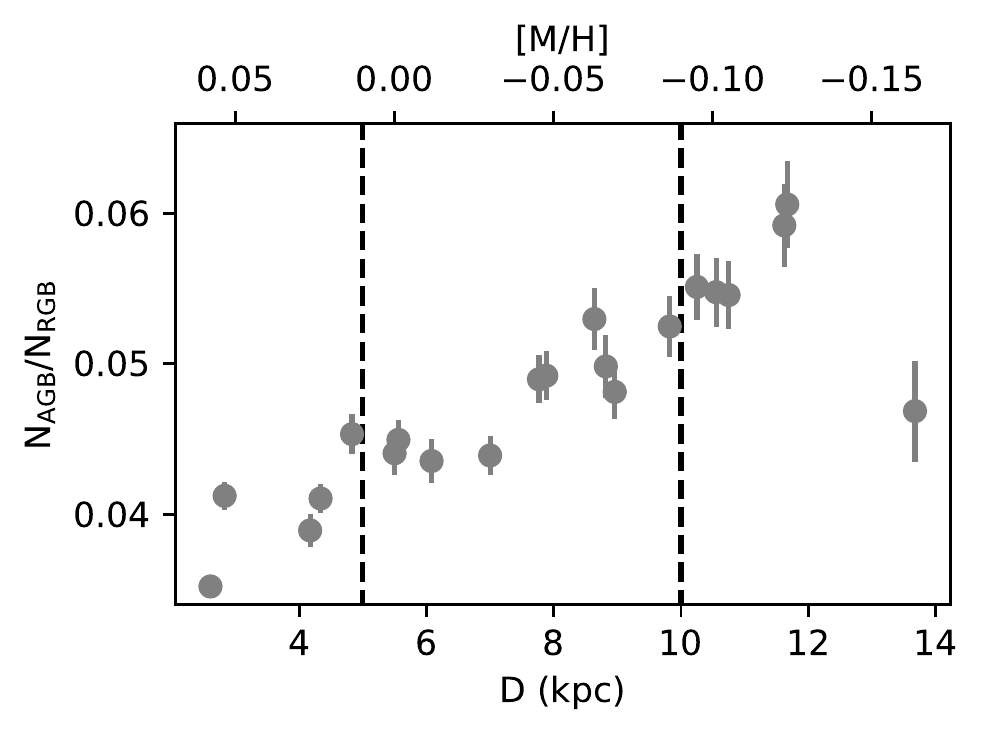}
  \end{center}
  
   \caption{$N_{AGB}/N_{RGB}$ as a population age proxy, plotted
     against deprojected distance. Lower values of $N_{\rm AGB}/N_{\rm
       RGB}$ indicate an older population. The 5 and 10 kpc
     star-forming rings are marked with dashed lines. As expected, the
     inner fields are generally older than the outer fields.}
  \label{fig:age}
\end{figure}

\begin{figure}
  \includegraphics[width=\columnwidth]{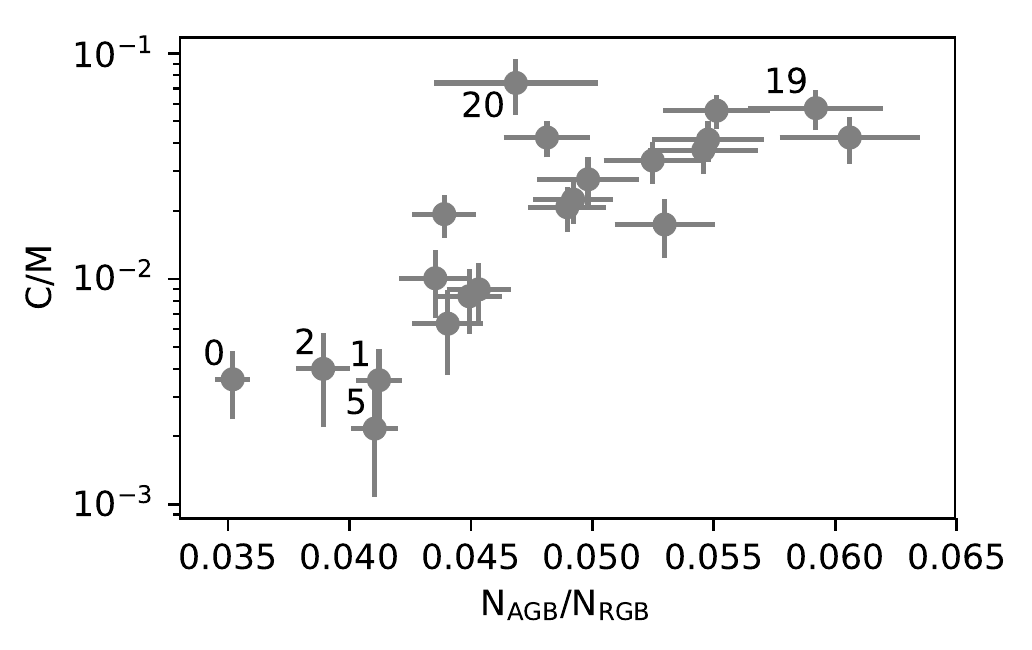}
  \caption{C/M vs. an age proxy ($N_{\rm AGB}/N_{\rm
      RGB}$), with error bars reflecting Poisson
    uncertainties. Both the age and metallicity are correlated with
    C/M, though metallicity shows the stronger correlation. In both
    panels, the lowest and highest metallicity fields are marked with
    the field number. }
  \label{fig:trend_age}
\end{figure}

While it appears that all of our fields include ages that produce
carbon stars, we can further assess the age of our fields and how it
affects C/M by calculating an age proxy: the ratio of TP-AGB stars to
RGB stars. This fraction is expected to be lower in populations with a
higher fraction of old stars because the TP-AGB phase has a limited
lifespan \citep[$\lesssim$3~Myr, e.g.,][]{Kalirai+2014, Cristallo+2015}. We define
$N_{\rm AGB}$ as the combined number of M and C stars, and $N_{\rm
  RGB}$ as the number of stars below the TRGB with $F814W-F110W > 0.5$
to avoid young main sequence stars and $F110W<22$~mag to avoid issues
with crowding and incompleteness. The resulting age proxy is plotted
against deprojected radius in Figure~\ref{fig:age}. As expected, the
innermost fields are older than the outer fields, on average. We also
see a hint of a younger population in the 5~kpc star-forming ring.

This age proxy is plotted against C/M in the righthand
  panel of Figure~\ref{fig:trend_age}. Both age and metallicity show a
strong correlation with C/M in our fields, which was also shown in the
outer regions of M31 by \citet{Hamren+2015}. The correlation matrix
derived using a Spearman rank correlation coefficient shows that the
degree of correlation between C/M and [M/H] is higher (96\%) than
between C/M and age (82\%). Therefore, while age is clearly important,
metallicity is the stronger effect on C/M in M31. A detailed analysis
that takes M31's star-formation history into account is being pursued
in a forthcoming paper (Chen et al., in preparation).

\subsubsection{Ruling out Differences in C/M Measurement Techniques}
\label{sec:check}

One could be concerned that the change in slope in
Figure~\ref{fig:lgcmr} is due to differences in classification
strategies. In contrast to our near-IR HST technique, most carbon star
surveys in the Local Group use optical narrow-band filters centered on
CN and TiO bands to identify carbon stars, including data from
\citet{Brewer+1995}, \citet{Battinelli+2003} and
\citet{Groenewegen2006}. We must therefore verify that our HST
medium-band data can be directly compared to the optical CN/TiO data.

The major difference among these strategies is the identification of M
giant stars, as noted in Section~\ref{sec:mstars}. Another notable
distinction lies in the fact that the optical data are not sensitive
to stars with a high amount of optical extinction from circumstellar
dust, which pushes both M-giants and C stars below the TRGB. However,
stars dusty enough to veil molecular features are rare
\citep[typically $<$5\%;][]{Boyer+2011,Srinivasan+2016,Boyer+2017}, so
the effect on C/M is expected to be small.

Despite these differences, C/M measured using both near-IR and optical
techniques agree very well in metal-poor galaxies (red circles and
cyan triangles in Fig.~\ref{fig:lgcmr}). This agreement is due to a
lack of late-type M giants in metal-poor galaxies and to weaker TiO
absorption in metal-poor stars at a given $T_{\rm eff}$, minimizing
the effect of the M star definition on C/M.  Metal-poor M giants fall
along a warmer Hayashi line \citep{Marigo+2013}, rarely reaching
spectral types later than approximately M3, though a few exceptions
exist in the Magellanic Clouds
\citep{Marshall+2004,vanLoon+2005,Goldman+2017}. Since only late-type
M giants fall below the TRGB in Figure~\ref{fig:mdef}, the C/M ratio
remains largely unaffected by details of choices of M star selections
at low metallicity.

In contrast, the disagreement in C/M grows at the high metallicity
end, resulting in the different values measured by different surveys
in M31 (note especially the offsets between this work,
\citet{Brewer+1995}, and \citet{Hamren+2015} in Fig.~\ref{fig:lgcmr}
at $[{\rm M/H}]>-0.25$). Nevertheless, both techniques indicate a
shift to a steeper C/M slope in M31 compared to metal-poor galaxies,
an effect that was not noted in the previous M31 studies despite,
e.g., the \citet{Brewer+1995} data showing a shift in slope in their 2
innermost fields. We therefore do not believe that the change in slope
in M31 is due to sample selection.

\subsection{Constraining Carbon Star Formation Efficiency}

As already noted, TP-AGB models predict a maximum metallicity limit
for carbon star formation. The exact limit is dictated by uncalibrated
model details, and is caused by a higher oxygen abundance,
less-efficient dredge up, and a late onset of the third
dredge-up. These characteristics prevent atmospheric C/O from
exceeding unity before mass loss terminates the TP-AGB phase
\citep{Karakas+2002,Marigo+2013}. Our results show that the
metallicity ceiling is a gradual shift, rather than an abrupt
  limit. An inflection point in carbon star formation efficiency, as
  indicated by the change in C/M slope in Figure~\ref{fig:lgcmr},
falls between the metallicities of the LMC and M31's disk, suggesting
that M33 could be an interesting environment in which to probe this
transition, given its intermediate metallicity.  The
\citet{Brewer+1995} and \citet{Battinelli+2003} data, and perhaps even
Field 20, already show a hint that this transition occurs in the
outermost region of M31's disk, possibly between the 10~kpc and 15~kpc
rings (Fig.~\ref{fig:map}).

On the high metallicity end, a hard limit for carbon star formation
remains elusive. Observations nearer to M31's bulge are likely the
best chance at identifying a limit since more distant galaxies suffer
from severe crowding and since the motion of large convective cells on
the stellar surface result in uncertain Gaia distance estimates to
TP-AGB stars in the Milky Way \citep{Chiavassa+2011, Chiavassa+2018}.

\section{Conclusions}
\label{sec:conc}

We have surveyed 20 fields in the disk of M31 ($R = 2.6$--$13.7$~kpc)
with HST's WFC3/IR camera to identify and classify TP-AGB stars. We
use the medium-band filters described in more detail in
\citet{Boyer+2013,Boyer+2017} to separate carbon-rich (C) from oxygen-rich
(M) TP-AGB stars and find that the ratio of the two (C/M) decreases
rapidly with galactic radius, resulting in a C/M that is more
than a factor of 10 lower than predicted by observations in more
metal-poor galaxies. The shift to a more rapid decline in C/M with
[M/H] appears to occur just beyond the 10~kpc star-forming ring,
corresponding to ${\rm [M/H]} > -0.1$~dex when assuming the
metallicity gradient derived by \citet{Gregersen+2015}. Despite this
rapid decline in C/M, carbon stars are still (inefficiently) forming
in M31 even at 2.6~kpc (${\rm [M/H]} \sim +0.06$). These observations
provide a robust constraint to models of TP-AGB stars, though the
predicted metallicity ceiling for carbon star formation remains
elusive.

\acknowledgements Support for HST program GO-14072 was provided by
NASA through a grant from the Space Telescope Science Institute, which
is operated by the Association of Universities for Research in
Astronomy, Incorporated, under NASA contract NAS5-26555. PM, LG, BA,
and YC acknowledge the support from the ERC Consolidator Grant scheme
(project STARKEY, G. A. n 615604). PR acknowledges support from the National
Science Foundation under Award No. 1501205.  DRW is supported by a
fellowship from the Alfred P. Sloan Foundation and acknowledges
support from the Alexander von Humboldt Foundation.

\bibliographystyle{aasjournal.bst}
\bibliography{myrefs}

\begin{thebibliography}{}
\expandafter\ifx\csname natexlab\endcsname\relax\def\natexlab#1{#1}\fi

\bibitem[{{Aringer} {et~al.}(2016){Aringer}, {Girardi}, {Nowotny}, {Marigo}, \&
  {Bressan}}]{Aringer+2016}
{Aringer}, B., {Girardi}, L., {Nowotny}, W., {Marigo}, P., \& {Bressan}, A.
  2016, \mnras, 457, 3611

\bibitem[{{Aringer} {et~al.}(2009){Aringer}, {Girardi}, {Nowotny}, {Marigo}, \&
  {Lederer}}]{Aringer+2009}
{Aringer}, B., {Girardi}, L., {Nowotny}, W., {Marigo}, P., \& {Lederer}, M.~T.
  2009, \aap, 503, 913

\bibitem[{Baldwin {et~al.}(2017)Baldwin, McDermid, Kuntschner, Maraston, \&
  Conroy}]{Baldwin+2017}
Baldwin, C., McDermid, R.~M., Kuntschner, H., Maraston, C., \& Conroy, C. 2017,
  Monthly Notices of the Royal Astronomical Society, 473, 4698

\bibitem[{{Barmby} {et~al.}(2006){Barmby}, {Ashby}, {Bianchi}, {Engelbracht},
  {Gehrz}, {Gordon}, {Hinz}, {Huchra}, {Humphreys}, {Pahre},
  {P{\'e}rez-Gonz{\'a}lez}, {Polomski}, {Rieke}, {Thilker}, {Willner}, \&
  {Woodward}}]{Barmby+2006}
{Barmby}, P., {Ashby}, M.~L.~N., {Bianchi}, L., {et~al.} 2006, \apjl, 650, L45

\bibitem[{{Battinelli} \& {Demers}(2005)}]{BattinelliDemers2005}
{Battinelli}, P., \& {Demers}, S. 2005, \aap, 434, 657

\bibitem[{{Battinelli} {et~al.}(2003){Battinelli}, {Demers}, \&
  {Letarte}}]{Battinelli+2003}
{Battinelli}, P., {Demers}, S., \& {Letarte}, B. 2003, \aj, 125, 1298

\bibitem[{{Blum} {et~al.}(2006){Blum}, {Mould}, {Olsen}, {Frogel}, {Werner},
  {Meixner}, {Markwick-Kemper}, {Indebetouw}, {Whitney}, {Meade}, {Babler},
  {Churchwell}, {Gordon}, {Engelbracht}, {For}, {Misselt}, {Vijh}, {Leitherer},
  {Volk}, {Points}, {Reach}, {Hora}, {Bernard}, {Boulanger}, {Bracker},
  {Cohen}, {Fukui}, {Gallagher}, {Gorjian}, {Harris}, {Kelly}, {Kawamura},
  {Latter}, {Madden}, {Mizuno}, {Mizuno}, {Nota}, {Oey}, {Onishi}, {Paladini},
  {Panagia}, {Perez-Gonzalez}, {Shibai}, {Sato}, {Smith}, {Staveley-Smith},
  {Tielens}, {Ueta}, {Van Dyk}, \& {Zaritsky}}]{Blum+2006}
{Blum}, R.~D., {Mould}, J.~R., {Olsen}, K.~A., {et~al.} 2006, \aj, 132, 2034

\bibitem[{{Boothroyd} \& {Sackmann}(1988)}]{BoothroydSackmann1988}
{Boothroyd}, A.~I., \& {Sackmann}, I.-J. 1988, \apj, 328, 632

\bibitem[{{Boyer} {et~al.}(2011){Boyer}, {Srinivasan}, {van Loon}, {McDonald},
  {Meixner}, {Zaritsky}, {Gordon}, {Kemper}, {Babler}, {Block}, {Bracker},
  {Engelbracht}, {Hora}, {Indebetouw}, {Meade}, {Misselt}, {Robitaille},
  {Sewi{\l}o}, {Shiao}, \& {Whitney}}]{Boyer+2011}
{Boyer}, M.~L., {Srinivasan}, S., {van Loon}, J.~{\relax Th}., {et~al.} 2011,
  \aj, 142, 103

\bibitem[{{Boyer} {et~al.}(2012){Boyer}, {Srinivasan}, {Riebel}, {McDonald},
  {van Loon}, {Clayton}, {Gordon}, {Meixner}, {Sargent}, \&
  {Sloan}}]{Boyer+2012}
{Boyer}, M.~L., {Srinivasan}, S., {Riebel}, D., {et~al.} 2012, \apj, 748, 40

\bibitem[{{Boyer} {et~al.}(2013){Boyer}, {Girardi}, {Marigo}, {Williams},
  {Aringer}, {Nowotny}, {Rosenfield}, {Dorman}, {Guhathakurta}, {Dalcanton},
  {Melbourne}, {Olsen}, \& {Weisz}}]{Boyer+2013}
{Boyer}, M.~L., {Girardi}, L., {Marigo}, P., {et~al.} 2013, \apj, 774, 83

\bibitem[{{Boyer} {et~al.}(2017){Boyer}, {McQuinn}, {Groenewegen}, {Zijlstra},
  {Whitelock}, {van Loon}, {Sonneborn}, {Sloan}, {Skillman}, {Meixner},
  {McDonald}, {Jones}, {Javadi}, {Gehrz}, {Britavskiy}, \&
  {Bonanos}}]{Boyer+2017}
{Boyer}, M.~L., {McQuinn}, K.~B.~W., {Groenewegen}, M.~A.~T., {et~al.} 2017,
  \apj, 851, 152

\bibitem[{{Bressan} {et~al.}(2012){Bressan}, {Marigo}, {Girardi}, {Salasnich},
  {Dal Cero}, {Rubele}, \& {Nanni}}]{Bressan+2012}
{Bressan}, A., {Marigo}, P., {Girardi}, L., {et~al.} 2012, \mnras, 427, 127

\bibitem[{Bressan {et~al.}(2014)Bressan, Tang, Slemer, Marigo, Rosenfield,
  Girardi, \& Bianchi}]{Tang+2014}
Bressan, A., Tang, J., Slemer, A., {et~al.} 2014, Monthly Notices of the Royal
  Astronomical Society, 445, 4287

\bibitem[{{Brewer} {et~al.}(1995){Brewer}, {Richer}, \&
  {Crabtree}}]{Brewer+1995}
{Brewer}, J.~P., {Richer}, H.~B., \& {Crabtree}, D.~R. 1995, \aj, 109, 2480

\bibitem[{{Cardelli} {et~al.}(1989){Cardelli}, {Clayton}, \&
  {Mathis}}]{Cardelli+1989}
{Cardelli}, J.~A., {Clayton}, G.~C., \& {Mathis}, J.~S. 1989, \apj, 345, 245

\bibitem[{{Chen} {et~al.}(2014){Chen}, {Girardi}, {Bressan}, {Marigo},
  {Barbieri}, \& {Kong}}]{Chen+2014}
{Chen}, Y., {Girardi}, L., {Bressan}, A., {et~al.} 2014, \mnras, 444, 2525

\bibitem[{{Chiavassa} {et~al.}(2018){Chiavassa}, {Freytag}, \&
  {Schultheis}}]{Chiavassa+2018}
{Chiavassa}, A., {Freytag}, B., \& {Schultheis}, M. 2018, \aap, 617, L1

\bibitem[{{Chiavassa} {et~al.}(2011){Chiavassa}, {Pasquato}, {Jorissen},
  {Sacuto}, {Babusiaux}, {Freytag}, {Ludwig}, {Cruzal{\`e}bes}, {Rabbia},
  {Spang}, \& {Chesneau}}]{Chiavassa+2011}
{Chiavassa}, A., {Pasquato}, E., {Jorissen}, A., {et~al.} 2011, \aap, 528, A120

\bibitem[{{Choi} {et~al.}(2016){Choi}, {Dotter}, {Conroy}, {Cantiello},
  {Paxton}, \& {Johnson}}]{Choi+2016}
{Choi}, J., {Dotter}, A., {Conroy}, C., {et~al.} 2016, \apj, 823, 102

\bibitem[{{Cioni}(2009)}]{Cioni2009}
{Cioni}, M.-R.~L. 2009, \aap, 506, 1137

\bibitem[{{Conroy} {et~al.}(2009){Conroy}, {Gunn}, \& {White}}]{Conroy+2009}
{Conroy}, C., {Gunn}, J.~E., \& {White}, M. 2009, \apj, 699, 486

\bibitem[{{Cristallo} {et~al.}(2015){Cristallo}, {Straniero}, {Piersanti}, \&
  {Gobrecht}}]{Cristallo+2015}
{Cristallo}, S., {Straniero}, O., {Piersanti}, L., \& {Gobrecht}, D. 2015,
  \apjs, 219, 40

\bibitem[{{Cristallo} {et~al.}(2011){Cristallo}, {Piersanti}, {Straniero},
  {Gallino}, {Dom{\'{\i}}nguez}, {Abia}, {Di Rico}, {Quintini}, \&
  {Bisterzo}}]{Cristallo+2011}
{Cristallo}, S., {Piersanti}, L., {Straniero}, O., {et~al.} 2011, \apjs, 197,
  17

\bibitem[{{Dalcanton} {et~al.}(2012{\natexlab{a}}){Dalcanton}, {Williams},
  {Melbourne}, {Girardi}, {Dolphin}, {Rosenfield}, {Boyer}, {de Jong},
  {Gilbert}, {Marigo}, {Olsen}, {Seth}, \& {Skillman}}]{Dalcanton+2012a}
{Dalcanton}, J.~J., {Williams}, B.~F., {Melbourne}, J.~L., {et~al.}
  2012{\natexlab{a}}, \apjs, 198, 6

\bibitem[{{Dalcanton} {et~al.}(2012{\natexlab{b}}){Dalcanton}, {Williams},
  {Lang}, {Lauer}, {Kalirai}, {Seth}, {Dolphin}, {Rosenfield}, {Weisz}, {Bell},
  {Bianchi}, {Boyer}, {Caldwell}, {Dong}, {Dorman}, {Gilbert}, {Girardi},
  {Gogarten}, {Gordon}, {Guhathakurta}, {Hodge}, {Holtzman}, {Johnson},
  {Larsen}, {Lewis}, {Melbourne}, {Olsen}, {Rix}, {Rosema}, {Saha},
  {Sarajedini}, {Skillman}, \& {Stanek}}]{Dalcanton+2012b}
{Dalcanton}, J.~J., {Williams}, B.~F., {Lang}, D., {et~al.} 2012{\natexlab{b}},
  \apjs, 200, 18

\bibitem[{{Dolphin}(2000)}]{Dolphin2000}
{Dolphin}, A.~E. 2000, \pasp, 112, 1383

\bibitem[{{Dorman} {et~al.}(2012){Dorman}, {Guhathakurta}, {Fardal}, {Lang},
  {Geha}, {Howley}, {Kalirai}, {Bullock}, {Cuillandre}, {Dalcanton}, {Gilbert},
  {Seth}, {Tollerud}, {Williams}, \& {Yniguez}}]{Dorman+2012}
{Dorman}, C.~E., {Guhathakurta}, P., {Fardal}, M.~A., {et~al.} 2012, \apj, 752,
  147

\bibitem[{{Fishlock} {et~al.}(2014){Fishlock}, {Karakas}, {Lugaro}, \&
  {Yong}}]{Fishlock+2014}
{Fishlock}, C.~K., {Karakas}, A.~I., {Lugaro}, M., \& {Yong}, D. 2014, \apj,
  797, 44

\bibitem[{{Fluks} {et~al.}(1994){Fluks}, {Plez}, {The}, {de Winter},
  {Westerlund}, \& {Steenman}}]{Fluks+1994}
{Fluks}, M.~A., {Plez}, B., {The}, P.~S., {et~al.} 1994, \aaps, 105, 311

\bibitem[{{Girardi} {et~al.}(2008){Girardi}, {Dalcanton}, {Williams}, {de
  Jong}, {Gallart}, {Monelli}, {Groenewegen}, {Holtzman}, {Olsen}, {Seth},
  {Weisz}, \& {the ANGST/ANGRRR Collaboration}}]{Girardi+2008}
{Girardi}, L., {Dalcanton}, J., {Williams}, B., {et~al.} 2008, \pasp, 120, 583

\bibitem[{{Goldman} {et~al.}(2017){Goldman}, {van Loon}, {Zijlstra}, {Green},
  {Wood}, {Nanni}, {Imai}, {Whitelock}, {Matsuura}, {Groenewegen}, \&
  {G{\'o}mez}}]{Goldman+2017}
{Goldman}, S.~R., {van Loon}, J.~{\relax Th}., {Zijlstra}, A.~A., {et~al.}
  2017, \mnras, 465, 403

\bibitem[{{Gregersen} {et~al.}(2015){Gregersen}, {Seth}, {Williams}, {Lang},
  {Dalcanton}, {Girardi}, {Skillman}, {Bell}, {Dolphin}, {Fouesneau},
  {Guhathakurta}, {Hamren}, {Johnson}, {Kalirai}, {Lewis}, {Monachesi}, \&
  {Olsen}}]{Gregersen+2015}
{Gregersen}, D., {Seth}, A.~C., {Williams}, B.~F., {et~al.} 2015, \aj, 150, 189

\bibitem[{{Groenewegen}(2006)}]{Groenewegen2006}
{Groenewegen}, M.~A.~T. 2006, \aap, 448, 181

\bibitem[{{Hamren} {et~al.}(2015){Hamren}, {Rockosi}, {Guhathakurta}, {Boyer},
  {Smith}, {Dalcanton}, {Gregersen}, {Seth}, {Lewis}, {Williams}, {Toloba},
  {Girardi}, {Dorman}, {Gilbert}, \& {Weisz}}]{Hamren+2015}
{Hamren}, K.~M., {Rockosi}, C.~M., {Guhathakurta}, P., {et~al.} 2015, \apj,
  810, 60

\bibitem[{{Johnson} {et~al.}(2013){Johnson}, {Weisz}, {Dalcanton}, {Johnson},
  {Dale}, {Dolphin}, {Gil de Paz}, {Kennicutt}, {Lee}, {Skillman}, {Boquien},
  \& {Williams}}]{Johnson+2013}
{Johnson}, B.~D., {Weisz}, D.~R., {Dalcanton}, J.~J., {et~al.} 2013, \apj, 772,
  8

\bibitem[{{Kalirai} {et~al.}(2014){Kalirai}, {Marigo}, \&
  {Tremblay}}]{Kalirai+2014}
{Kalirai}, J.~S., {Marigo}, P., \& {Tremblay}, P.-E. 2014, \apj, 782, 17

\bibitem[{{Karakas}(2014)}]{Karakas2014}
{Karakas}, A.~I. 2014, \mnras, 445, 347

\bibitem[{{Karakas} \& {Lattanzio}(2014)}]{KarakasLattanzio2014}
{Karakas}, A.~I., \& {Lattanzio}, J.~C. 2014, \pasa, 31, e030

\bibitem[{{Karakas} {et~al.}(2002){Karakas}, {Lattanzio}, \&
  {Pols}}]{Karakas+2002}
{Karakas}, A.~I., {Lattanzio}, J.~C., \& {Pols}, O.~R. 2002, PASA, 19, 515

\bibitem[{{Karakas} \& {Lugaro}(2016)}]{Karakas+2016}
{Karakas}, A.~I., \& {Lugaro}, M. 2016, \apj, 825, 26

\bibitem[{{Karakas} {et~al.}(2018){Karakas}, {Lugaro}, {Carlos}, {Cseh},
  {Kamath}, \& {Garc{\'{\i}}a-Hern{\'a}ndez}}]{Karakas+2018}
{Karakas}, A.~I., {Lugaro}, M., {Carlos}, M., {et~al.} 2018, \mnras, 477, 421

\bibitem[{{Lewis} {et~al.}(2015){Lewis}, {Dolphin}, {Dalcanton}, {Weisz},
  {Williams}, {Bell}, {Seth}, {Simones}, {Skillman}, {Choi}, {Fouesneau},
  {Guhathakurta}, {Johnson}, {Kalirai}, {Leroy}, {Monachesi}, {Rix}, \&
  {Schruba}}]{Lewis+2015}
{Lewis}, A.~R., {Dolphin}, A.~E., {Dalcanton}, J.~J., {et~al.} 2015, \apj, 805,
  183

\bibitem[{{Marigo} {et~al.}(2013){Marigo}, {Bressan}, {Nanni}, {Girardi}, \&
  {Pumo}}]{Marigo+2013}
{Marigo}, P., {Bressan}, A., {Nanni}, A., {Girardi}, L., \& {Pumo}, M.~L. 2013,
  \mnras, 434, 488

\bibitem[{{Marshall} {et~al.}(2004){Marshall}, {van Loon}, {Matsuura}, {Wood},
  {Zijlstra}, \& {Whitelock}}]{Marshall+2004}
{Marshall}, J.~R., {van Loon}, J.~{\relax Th}., {Matsuura}, M., {et~al.} 2004,
  \mnras, 355, 1348

\bibitem[{{Matsuura} {et~al.}(2009){Matsuura}, {Barlow}, {Zijlstra},
  {Whitelock}, {Cioni}, {Groenewegen}, {Volk}, {Kemper}, {Kodama}, {Lagadec},
  {Meixner}, {Sloan}, \& {Srinivasan}}]{Matsuura+2009}
{Matsuura}, M., {Barlow}, M.~J., {Zijlstra}, A.~A., {et~al.} 2009, \mnras, 396,
  918

\bibitem[{{McConnachie} {et~al.}(2005){McConnachie}, {Irwin}, {Ferguson},
  {Ibata}, {Lewis}, \& {Tanvir}}]{McConnachie+2005}
{McConnachie}, A.~W., {Irwin}, M.~J., {Ferguson}, A.~M.~N., {et~al.} 2005,
  \mnras, 356, 979

\bibitem[{{McQuinn} {et~al.}(2019){McQuinn}, {Boyer}, {Skillman}, \&
  {Dolphin}}]{McQuinn+2019}
{McQuinn}, K.~B.~W., {Boyer}, M.~L., {Skillman}, E.~D., \& {Dolphin}, A. 2019,
  \apj, submitted

\bibitem[{{McQuinn} {et~al.}(2017){McQuinn}, {Boyer}, {Mitchell}, {Skillman},
  {Gehrz}, {Groenewegen}, {McDonald}, {Sloan}, {van Loon}, {Whitelock}, \&
  {Zijlstra}}]{McQuinn+2017}
{McQuinn}, K.~B.~W., {Boyer}, M.~L., {Mitchell}, M.~B., {et~al.} 2017, \apj,
  834, 78

\bibitem[{{Melbourne} \& {Boyer}(2013)}]{MelbourneBoyer2013}
{Melbourne}, J., \& {Boyer}, M.~L. 2013, \apj, 764, 30

\bibitem[{{Melbourne} {et~al.}(2012){Melbourne}, {Williams}, {Dalcanton},
  {Rosenfield}, {Girardi}, {Marigo}, {Weisz}, {Dolphin}, {Boyer}, {Olsen},
  {Skillman}, \& {Seth}}]{Melbourne+2012}
{Melbourne}, J., {Williams}, B.~F., {Dalcanton}, J.~J., {et~al.} 2012, \apj,
  748, 47

\bibitem[{{M{\'e}ndez} {et~al.}(2002){M{\'e}ndez}, {Davis}, {Moustakas},
  {Newman}, {Madore}, \& {Freedman}}]{Mendez+2002}
{M{\'e}ndez}, B., {Davis}, M., {Moustakas}, J., {et~al.} 2002, \aj, 124, 213

\bibitem[{{Nanni} {et~al.}(2013){Nanni}, {Bressan}, {Marigo}, \&
  {Girardi}}]{Nanni+2013}
{Nanni}, A., {Bressan}, A., {Marigo}, P., \& {Girardi}, L. 2013, \mnras, 434,
  2390

\bibitem[{{Nanni} {et~al.}(2016){Nanni}, {Marigo}, {Groenewegen}, {Aringer},
  {Girardi}, {Pastorelli}, {Bressan}, \& {Bladh}}]{Nanni+2016}
{Nanni}, A., {Marigo}, P., {Groenewegen}, M.~A.~T., {et~al.} 2016, \mnras, 462,
  1215

\bibitem[{{Nataf} {et~al.}(2011){Nataf}, {Udalski}, {Gould}, \&
  {Pinsonneault}}]{Nataf+2011}
{Nataf}, D.~M., {Udalski}, A., {Gould}, A., \& {Pinsonneault}, M.~H. 2011,
  \apj, 730, 118

\bibitem[{{Nowotny} {et~al.}(2001){Nowotny}, {Kerschbaum}, {Schwarz}, \&
  {Olofsson}}]{Nowotny+2001}
{Nowotny}, W., {Kerschbaum}, F., {Schwarz}, H.~E., \& {Olofsson}, H. 2001,
  \aap, 367, 557

\bibitem[{{Pastorelli} {et~al.}(2019){Pastorelli}, {Marigo}, {Girardi}, {Chen},
  {Rubele}, {Trabucchi}, {Aringer}, {Bladh}, \& {Bressan}}]{Pastorelli+2019}
{Pastorelli}, G., {Marigo}, P., {Girardi}, L., {et~al.} 2019, \mnras, submitted

\bibitem[{{Pietrinferni} {et~al.}(2007){Pietrinferni}, {Cassisi}, {Salaris},
  {Cordier}, \& {Castelli}}]{Pietrinferni+2007}
{Pietrinferni}, A., {Cassisi}, S., {Salaris}, M., {Cordier}, D., \& {Castelli},
  F. 2007, in IAU Symposium, Vol. 241, Stellar Populations as Building Blocks
  of Galaxies, ed. A.~{Vazdekis} \& R.~{Peletier}, 39--40

\bibitem[{{Rosenfield} {et~al.}(2016){Rosenfield}, {Marigo}, {Girardi},
  {Dalcanton}, {Bressan}, {Williams}, \& {Dolphin}}]{Rosenfield+2016}
{Rosenfield}, P., {Marigo}, P., {Girardi}, L., {et~al.} 2016, \apj, 822, 73

\bibitem[{{Rosenfield} {et~al.}(2012){Rosenfield}, {Johnson}, {Girardi},
  {Dalcanton}, {Bressan}, {Lang}, {Williams}, {Guhathakurta}, {Howley},
  {Lauer}, {Bell}, {Bianchi}, {Caldwell}, {Dolphin}, {Dorman}, {Gilbert},
  {Kalirai}, {Larsen}, {Olsen}, {Rix}, {Seth}, {Skillman}, \&
  {Weisz}}]{Rosenfield+2012}
{Rosenfield}, P., {Johnson}, L.~C., {Girardi}, L., {et~al.} 2012, \apj, 755,
  131

\bibitem[{{Rosenfield} {et~al.}(2014){Rosenfield}, {Marigo}, {Girardi},
  {Dalcanton}, {Bressan}, {Gullieuszik}, {Weisz}, {Williams}, {Dolphin}, \&
  {Aringer}}]{Rosenfield+2014}
{Rosenfield}, P., {Marigo}, P., {Girardi}, L., {et~al.} 2014, \apj, 790, 22

\bibitem[{{Saglia} {et~al.}(2018){Saglia}, {Opitsch}, {Fabricius}, {Bender},
  {Bla{\~n}a}, \& {Gerhard}}]{Saglia+2018}
{Saglia}, R.~P., {Opitsch}, M., {Fabricius}, M.~H., {et~al.} 2018, \aap, 618,
  A156

\bibitem[{{Sanders} {et~al.}(2012){Sanders}, {Caldwell}, {McDowell}, \&
  {Harding}}]{Sanders+2012}
{Sanders}, N.~E., {Caldwell}, N., {McDowell}, J., \& {Harding}, P. 2012, \apj,
  758, 133

\bibitem[{{Schlegel} {et~al.}(1998){Schlegel}, {Finkbeiner}, \&
  {Davis}}]{Schlegel+1998}
{Schlegel}, D.~J., {Finkbeiner}, D.~P., \& {Davis}, M. 1998, \apj, 500, 525

\bibitem[{{Schneider} {et~al.}(2014){Schneider}, {Valiante}, {Ventura},
  {dell'Agli}, {Di Criscienzo}, {Hirashita}, \& {Kemper}}]{Schneider+2014}
{Schneider}, R., {Valiante}, R., {Ventura}, P., {et~al.} 2014, \mnras, 442,
  1440

\bibitem[{{Sirianni} {et~al.}(2005){Sirianni}, {Jee}, {Ben{\'{\i}}tez},
  {Blakeslee}, {Martel}, {Meurer}, {Clampin}, {De Marchi}, {Ford}, {Gilliland},
  {Hartig}, {Illingworth}, {Mack}, \& {McCann}}]{Sirianni+2005}
{Sirianni}, M., {Jee}, M.~J., {Ben{\'{\i}}tez}, N., {et~al.} 2005, \pasp, 117,
  1049

\bibitem[{{Srinivasan} {et~al.}(2016){Srinivasan}, {Boyer}, {Kemper},
  {Meixner}, {Sargent}, \& {Riebel}}]{Srinivasan+2016}
{Srinivasan}, S., {Boyer}, M.~L., {Kemper}, F., {et~al.} 2016, \mnras, 457,
  2814

\bibitem[{{van Loon} {et~al.}(2005){van Loon}, {Cioni}, {Zijlstra}, \&
  {Loup}}]{vanLoon+2005}
{van Loon}, J.~{\relax Th}., {Cioni}, M.-R.~L., {Zijlstra}, A.~A., \& {Loup},
  C. 2005, \aap, 438, 273

\bibitem[{{Ventura} \& {Marigo}(2010)}]{VenturaMarigo2010}
{Ventura}, P., \& {Marigo}, P. 2010, \mnras, 408, 2476

\bibitem[{{Weiss} \& {Ferguson}(2009)}]{WeissFerguson2009}
{Weiss}, A., \& {Ferguson}, J.~W. 2009, \aap, 508, 1343

\bibitem[{{Williams} {et~al.}(2014){Williams}, {Lang}, {Dalcanton}, {Dolphin},
  {Weisz}, {Bell}, {Bianchi}, {Byler}, {Gilbert}, {Girardi}, {Gordon},
  {Gregersen}, {Johnson}, {Kalirai}, {Lauer}, {Monachesi}, {Rosenfield},
  {Seth}, \& {Skillman}}]{Williams+2014}
{Williams}, B.~F., {Lang}, D., {Dalcanton}, J.~J., {et~al.} 2014, \apjs, 215, 9

\bibitem[{{Williams} {et~al.}(2017){Williams}, {Dolphin}, {Dalcanton}, {Weisz},
  {Bell}, {Lewis}, {Rosenfield}, {Choi}, {Skillman}, \&
  {Monachesi}}]{Williams+2017}
{Williams}, B.~F., {Dolphin}, A.~E., {Dalcanton}, J.~J., {et~al.} 2017, \apj,
  846, 145

\bibitem[{{Zhukovska} \& {Henning}(2013)}]{Zhukovska+2013}
{Zhukovska}, S., \& {Henning}, T. 2013, \aap, 555, A99

\bibitem[{{Zurita} \& {Bresolin}(2012)}]{ZuritaBresolin2012}
{Zurita}, A., \& {Bresolin}, F. 2012, \mnras, 427, 1463

\end{thebibliography}

\newpage

\appendix

\section{IR Medium-band Color-color Diagrams}

We include here the color-color diagrams (CCDs) constructed from all 21
fields, including the pilot field (Field 0). Figure~\ref{fig:mainccd}
combines all fields into a single CCD.

\begin{figure}
  \hbox{
  \includegraphics[width=0.31\textwidth]{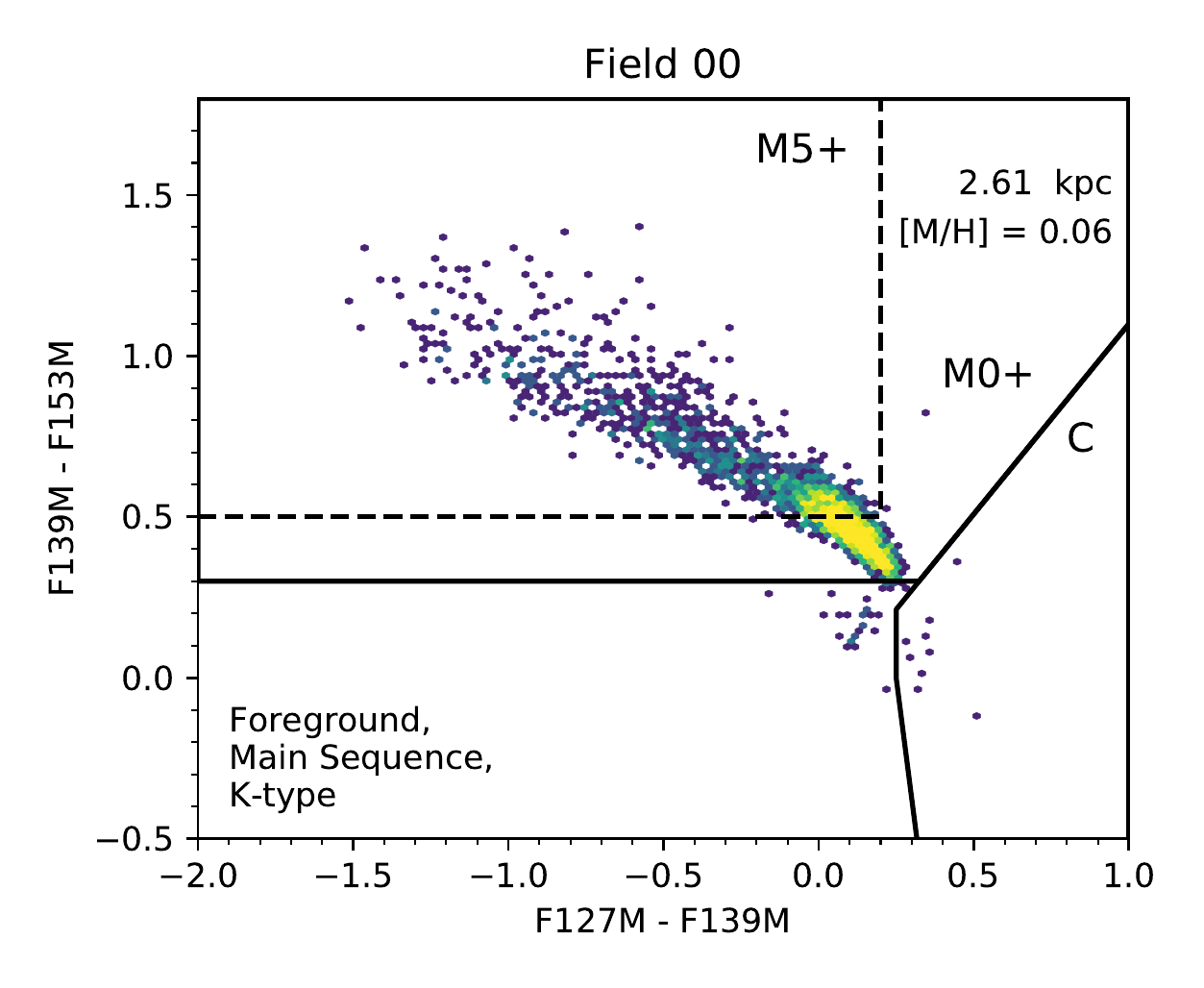}
  \includegraphics[width=0.31\textwidth]{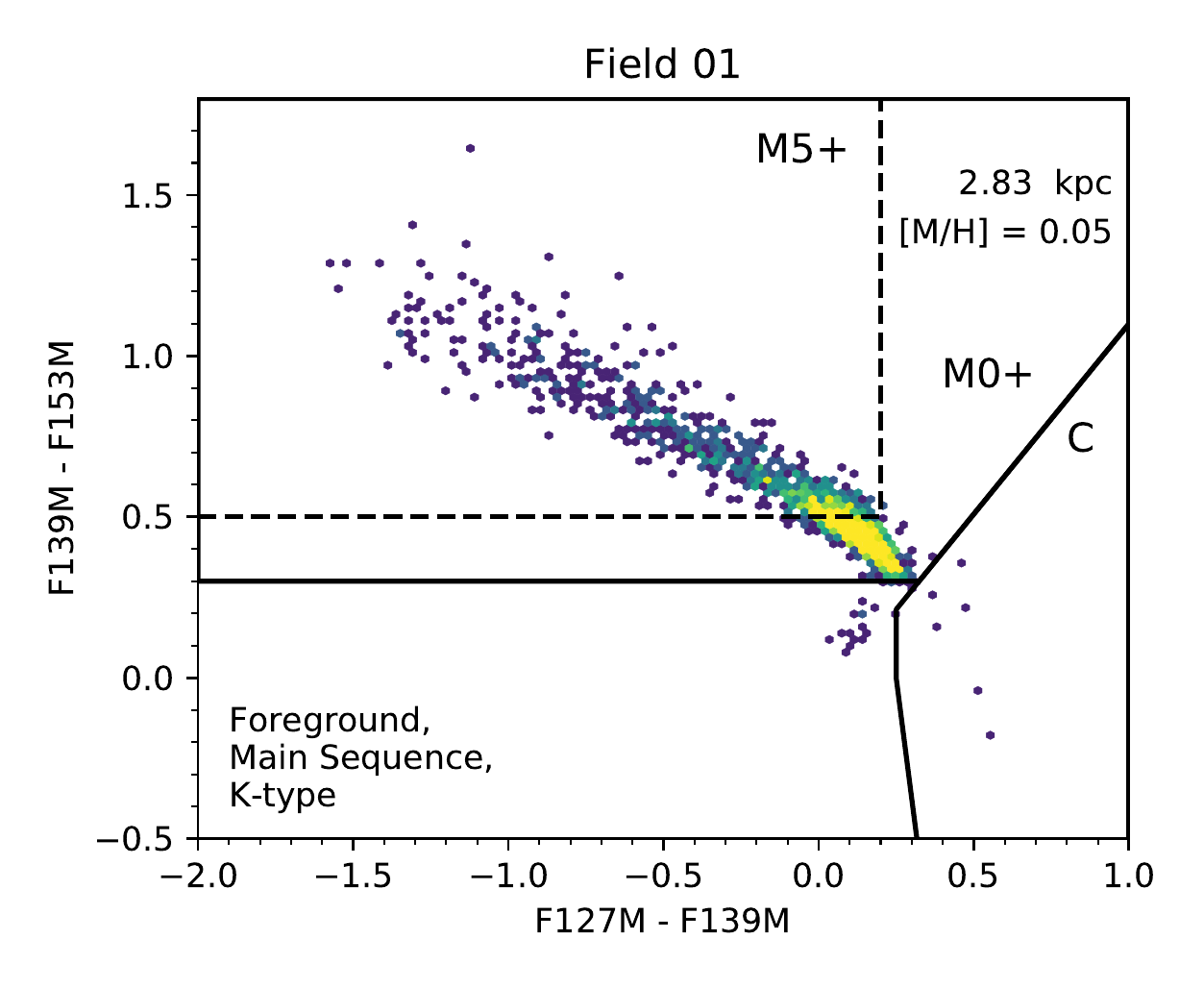}
  \includegraphics[width=0.31\textwidth]{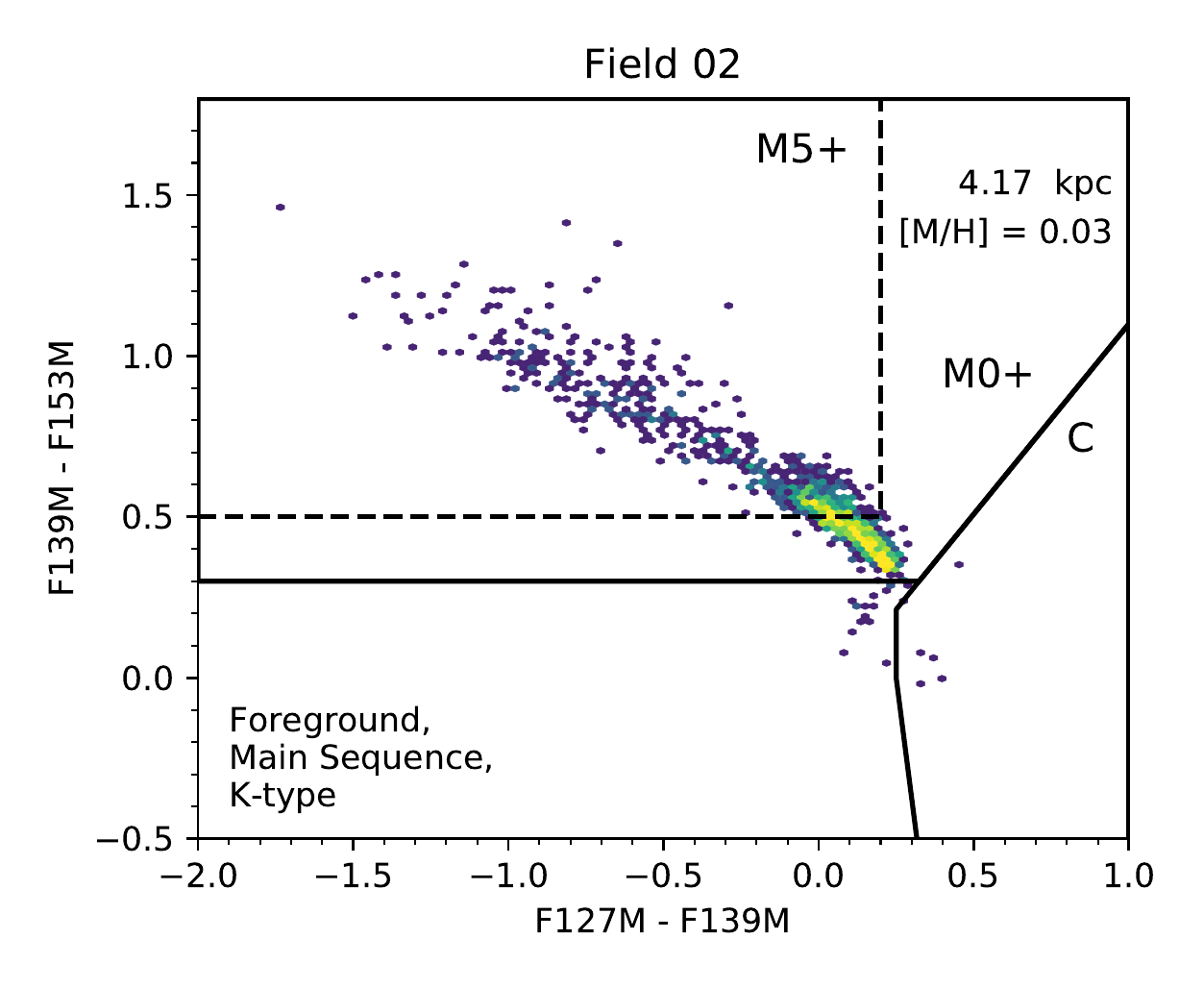}
}
  \hbox{
  \includegraphics[width=0.31\textwidth]{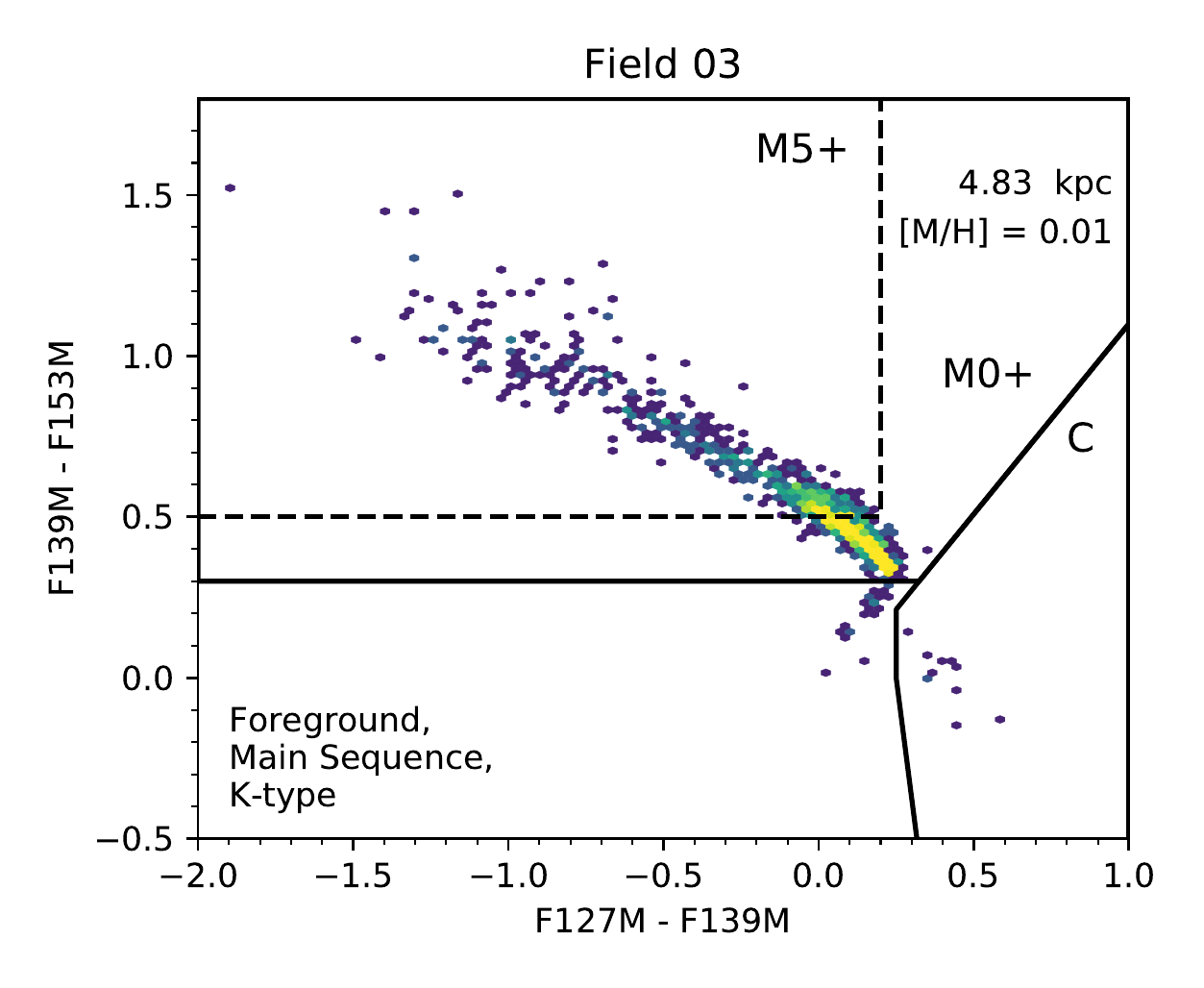}
  \includegraphics[width=0.31\textwidth]{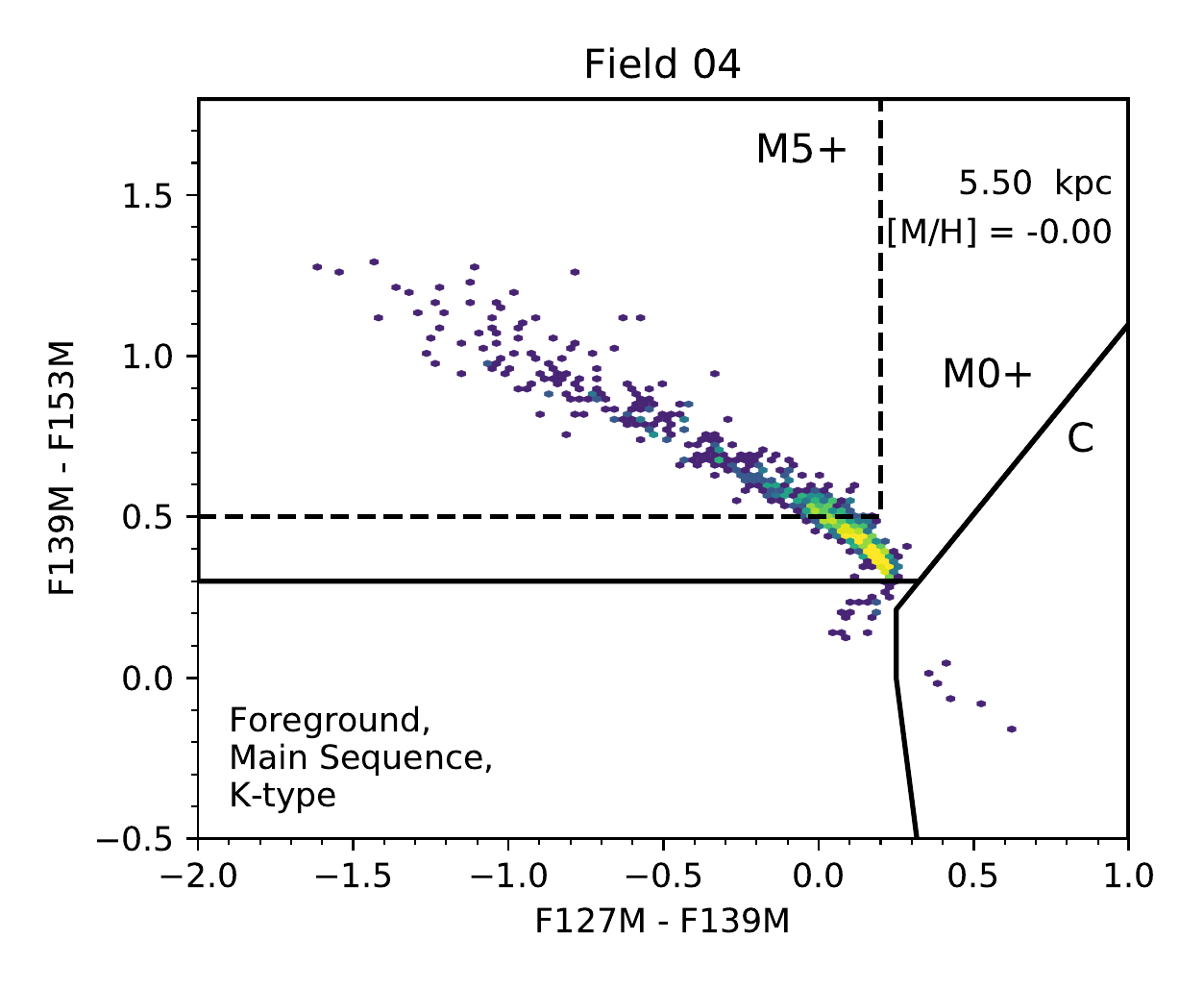}
  \includegraphics[width=0.31\textwidth]{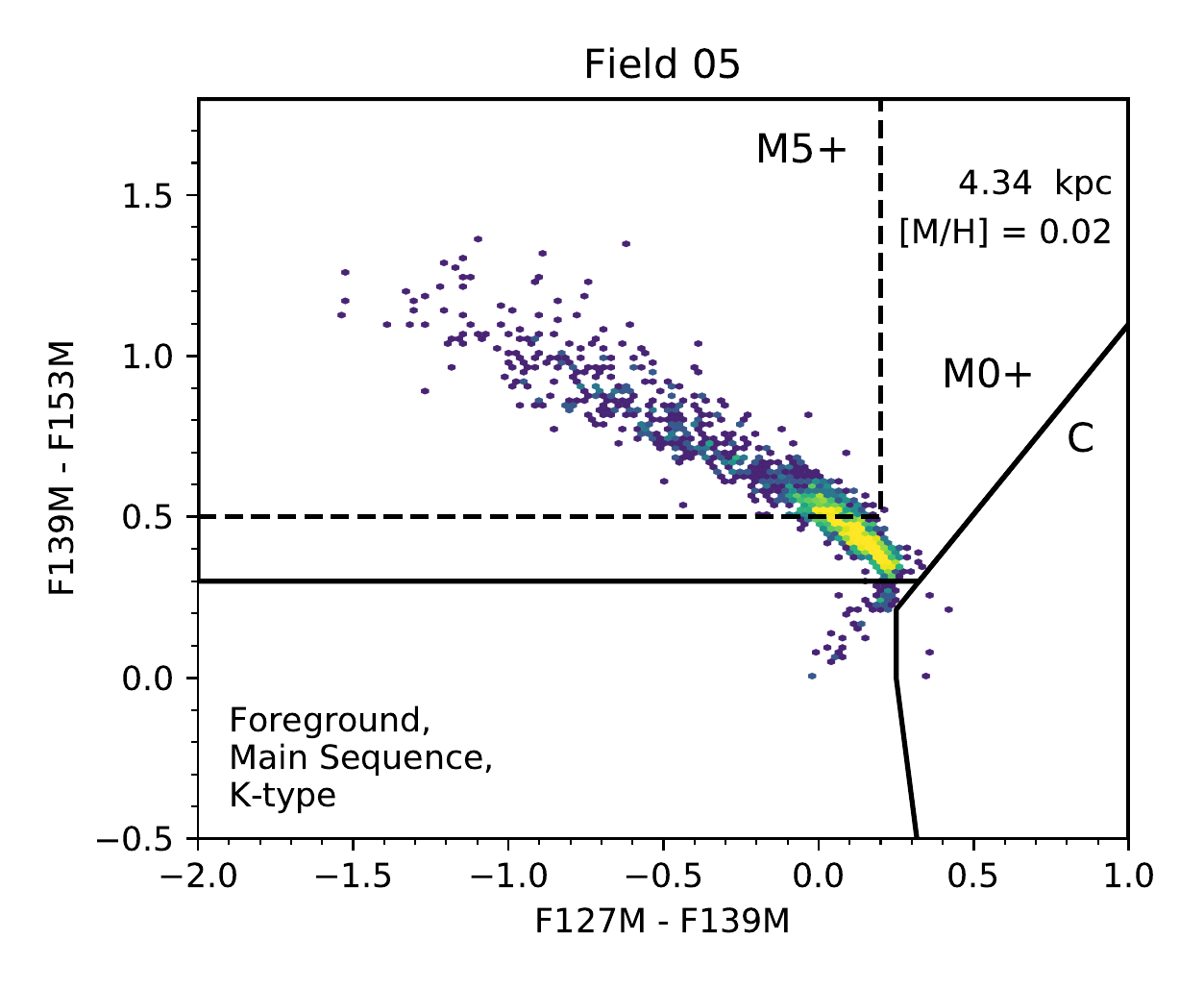}
  }                                     
  \hbox{                                
  \includegraphics[width=0.31\textwidth]{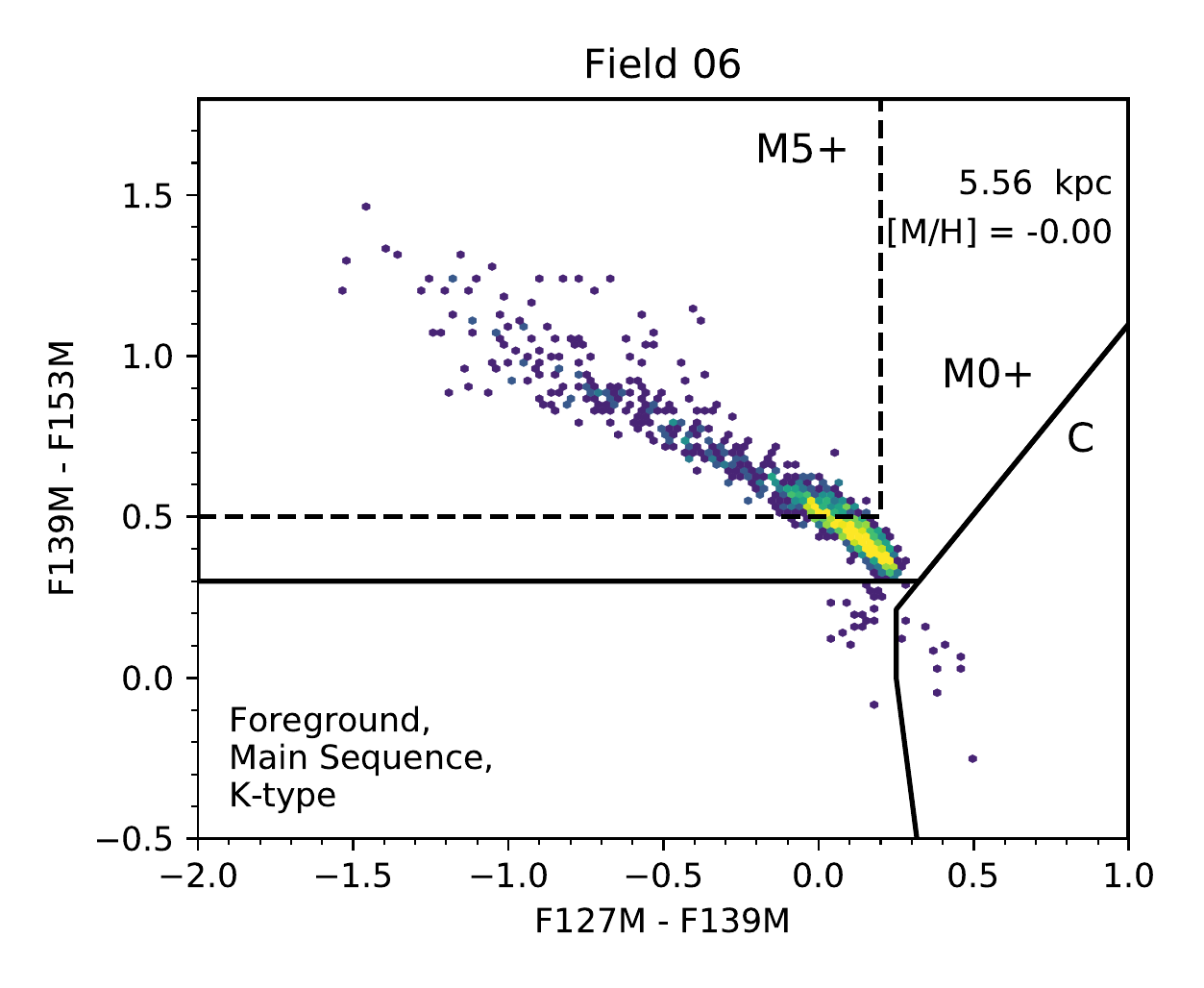}
  \includegraphics[width=0.31\textwidth]{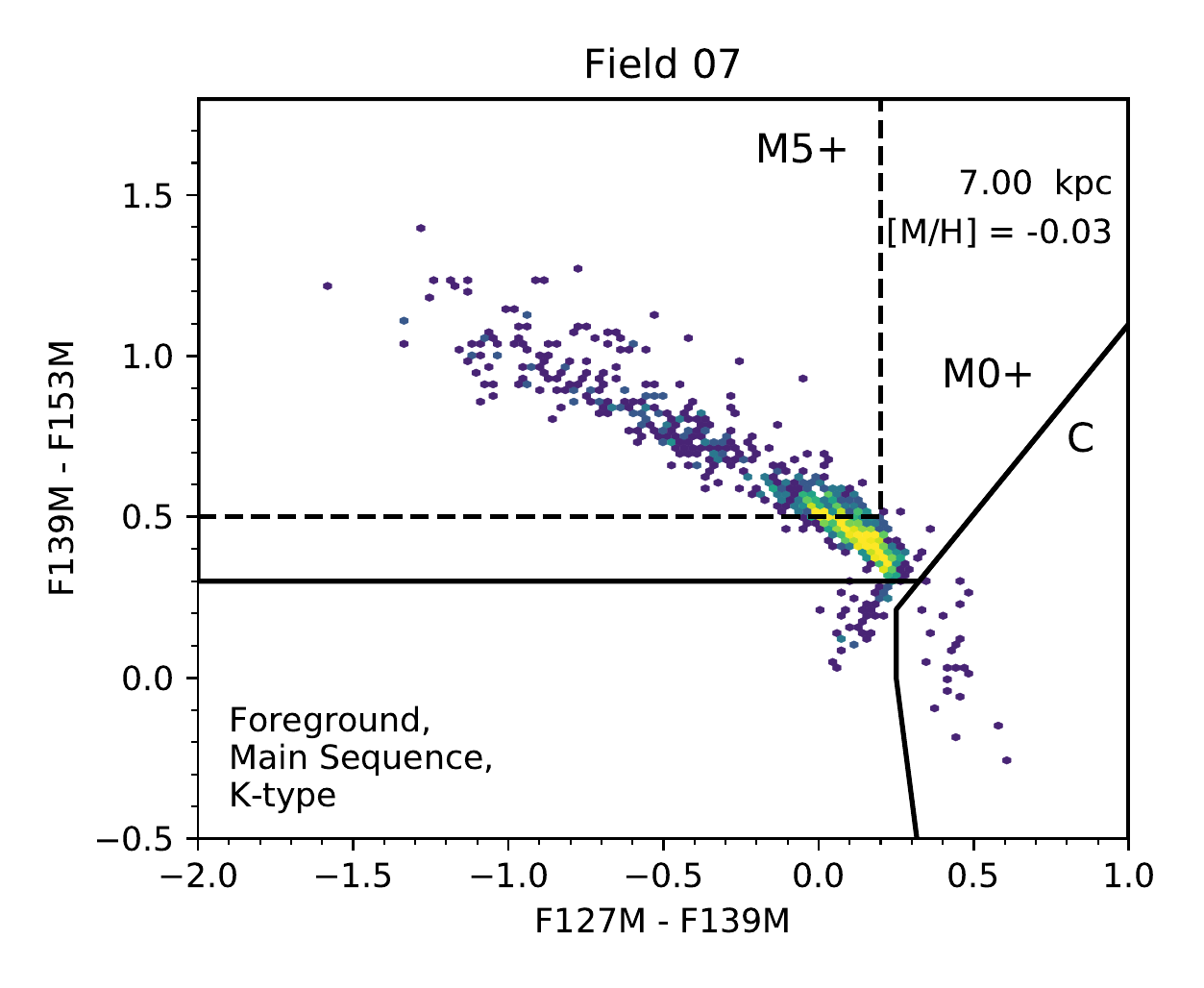}
  \includegraphics[width=0.31\textwidth]{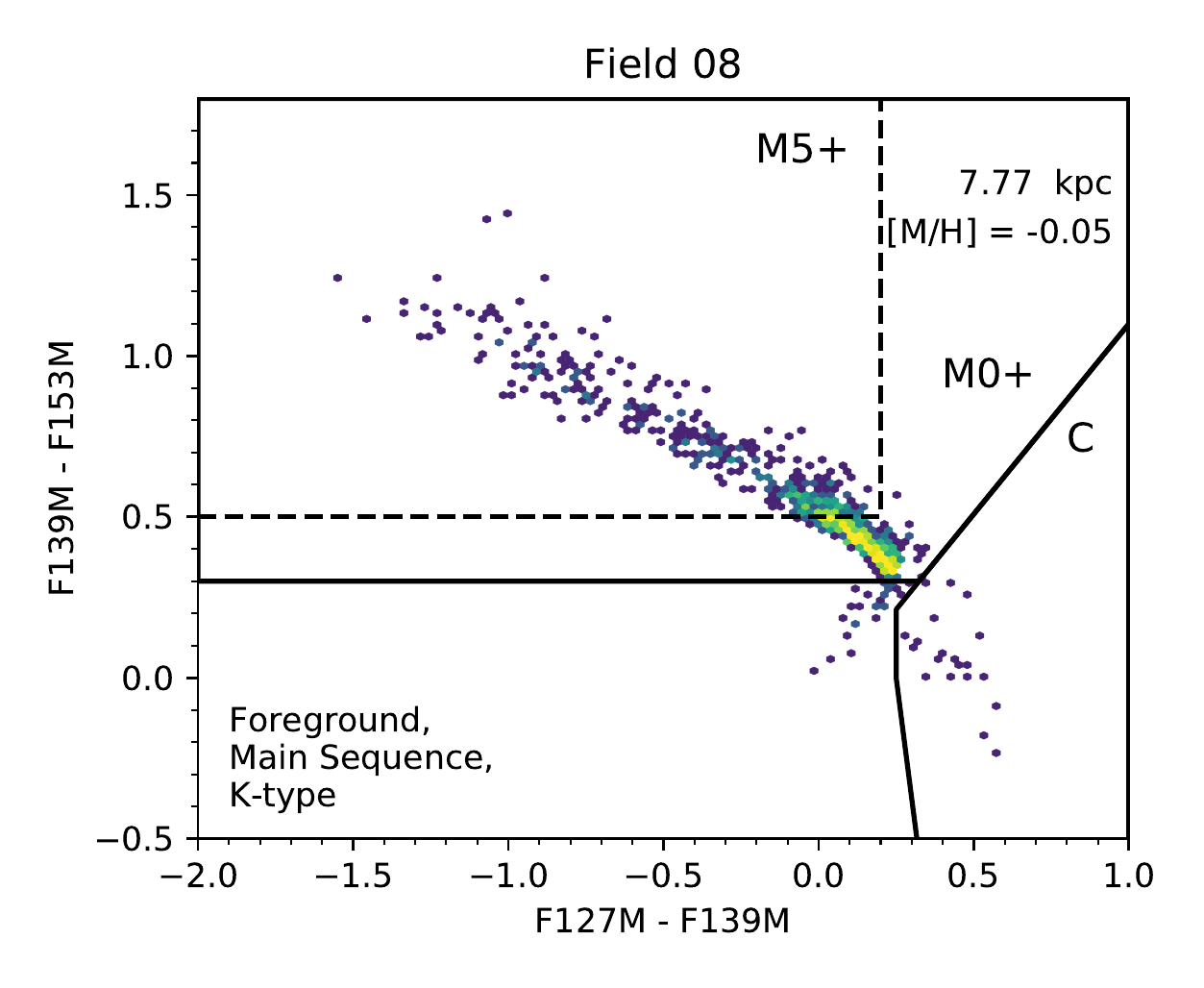}
  }                                     
  \hbox{                                
  \includegraphics[width=0.31\textwidth]{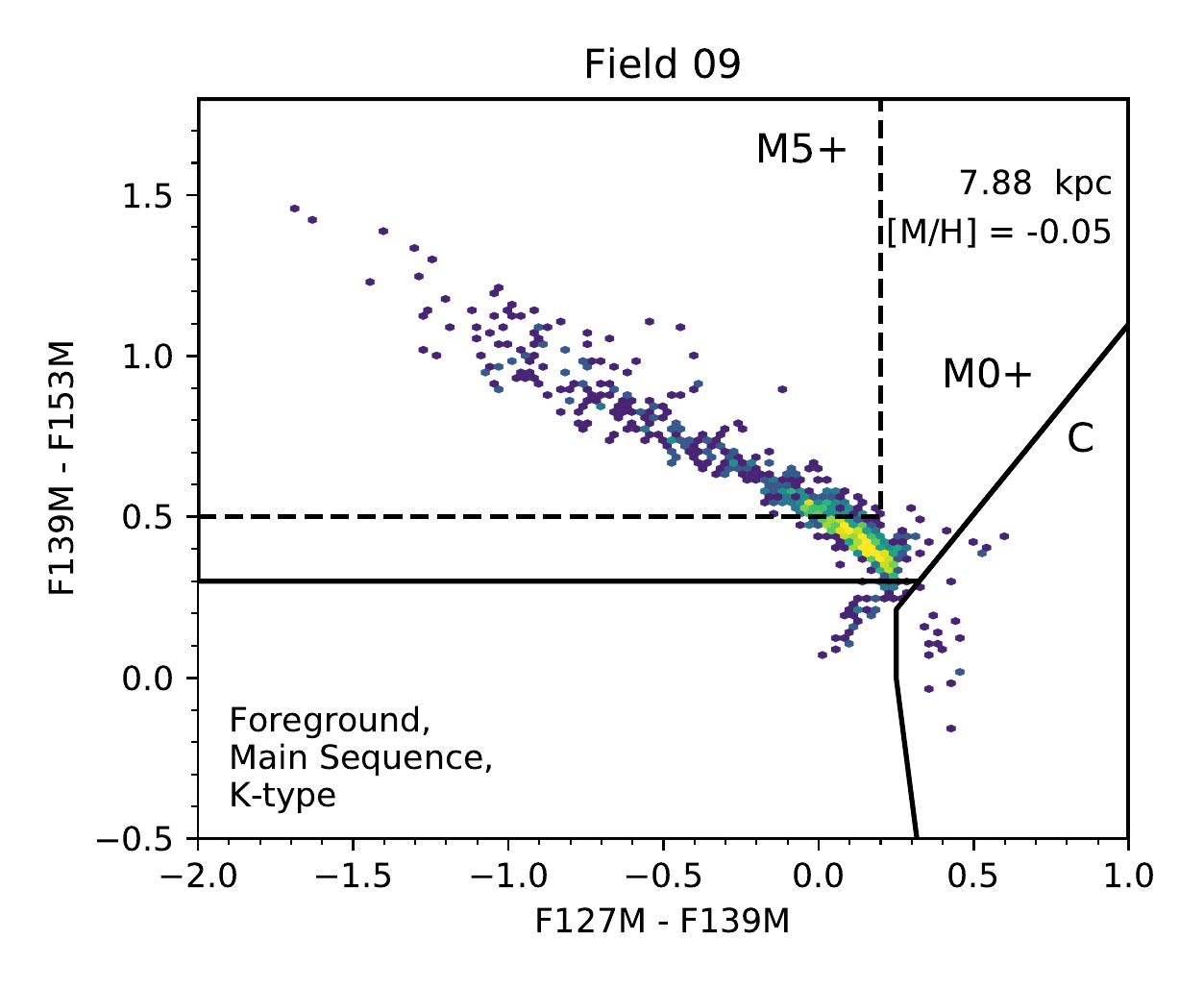}
  \includegraphics[width=0.31\textwidth]{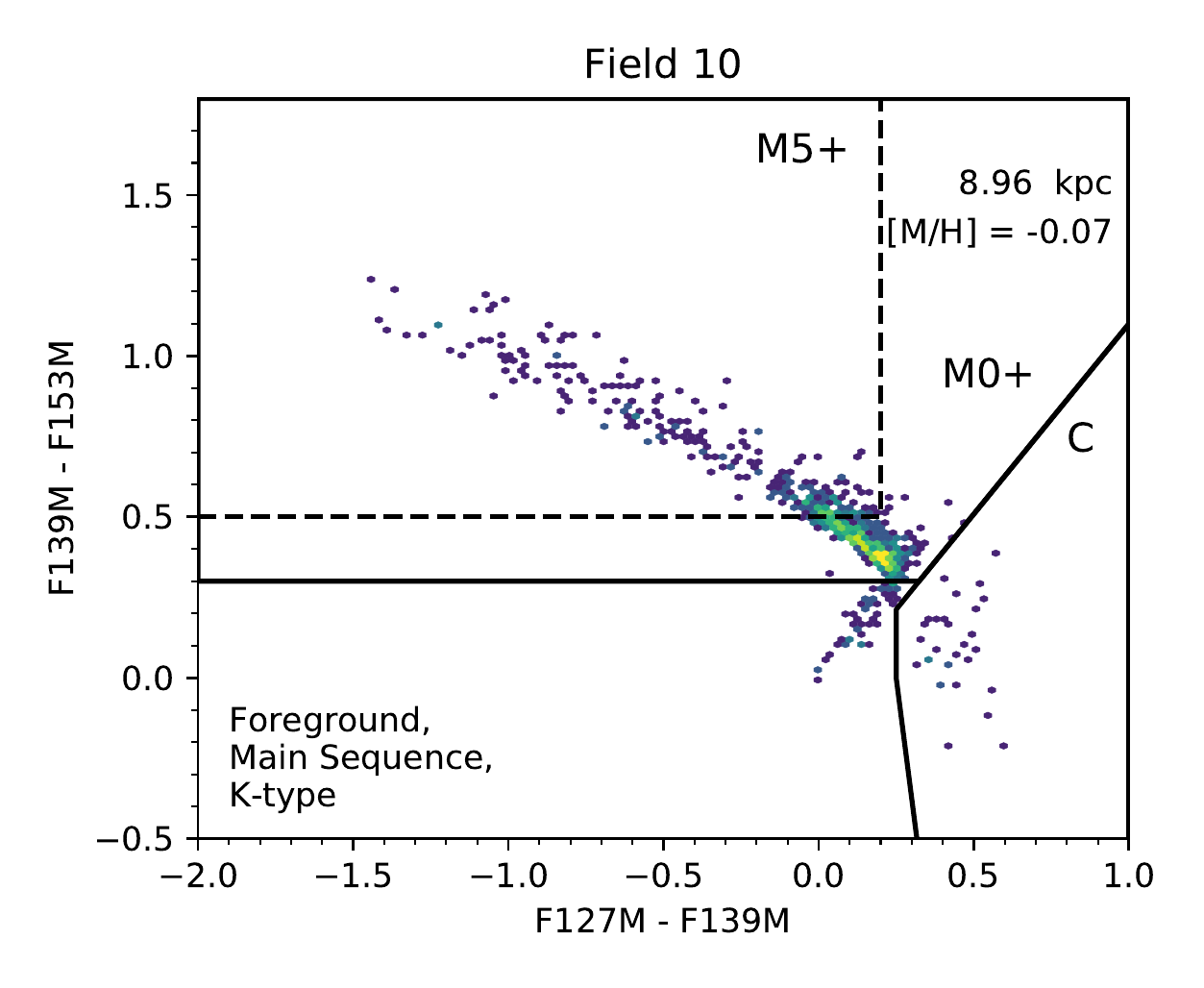}
  \includegraphics[width=0.31\textwidth]{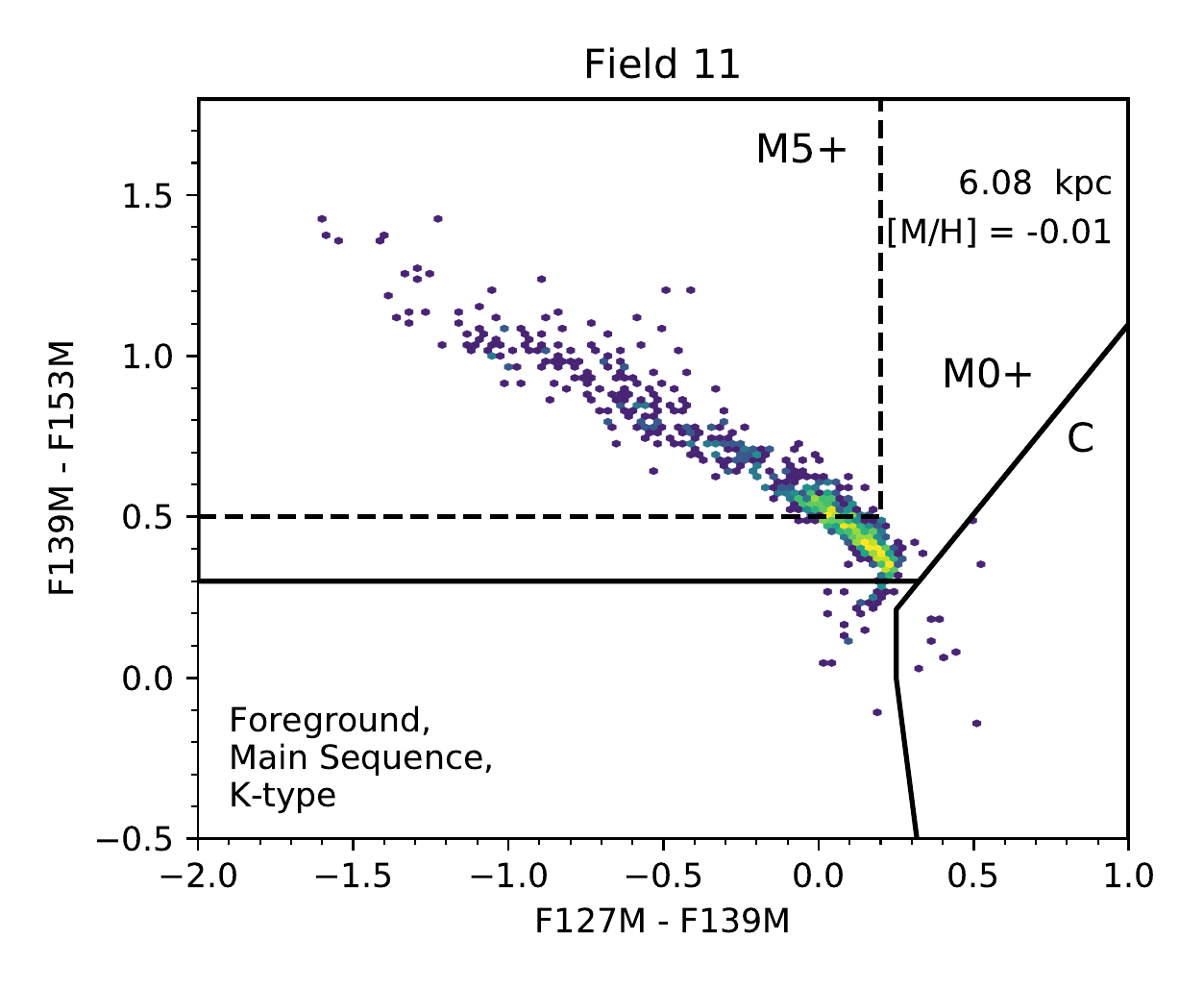}
  }

  \caption{Medium-band IR CCDs for Fields 0--11. As in
    Figure~\ref{fig:mainccd}, only stars brighter than one of the
    near-IR TRGBs are included. Solid lines separate M-type from
    C-type TP-AGB stars and other contaminating sources (Foreground,
    main sequence, and K-type). Stars to the left of the dashed line
    are late-type M giants (later than M5).}
  \label{fig:appendix_ccd1}
\end{figure}

\begin{figure}
  \hbox{
  \includegraphics[width=0.31\textwidth]{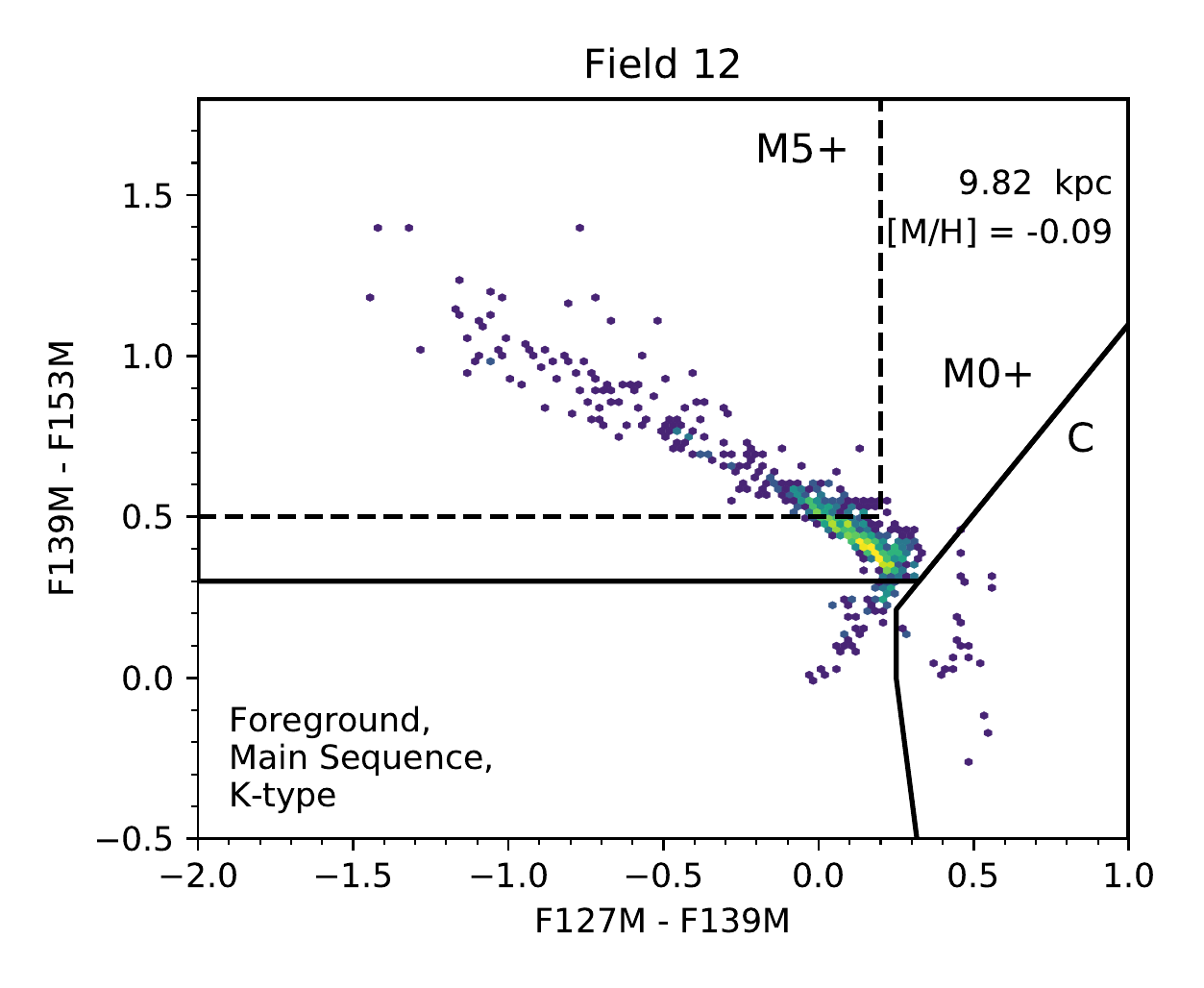}
  \includegraphics[width=0.31\textwidth]{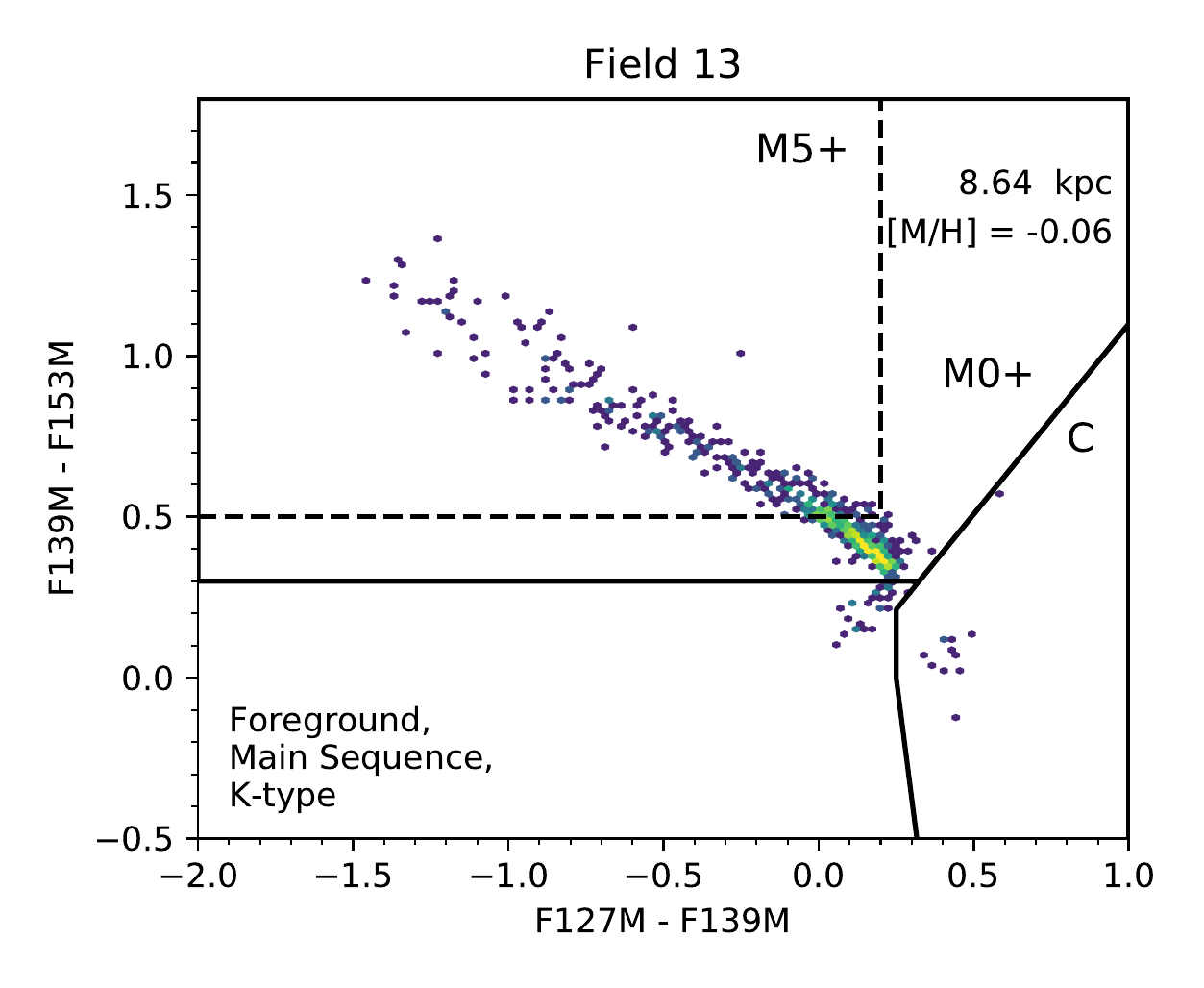}
  \includegraphics[width=0.31\textwidth]{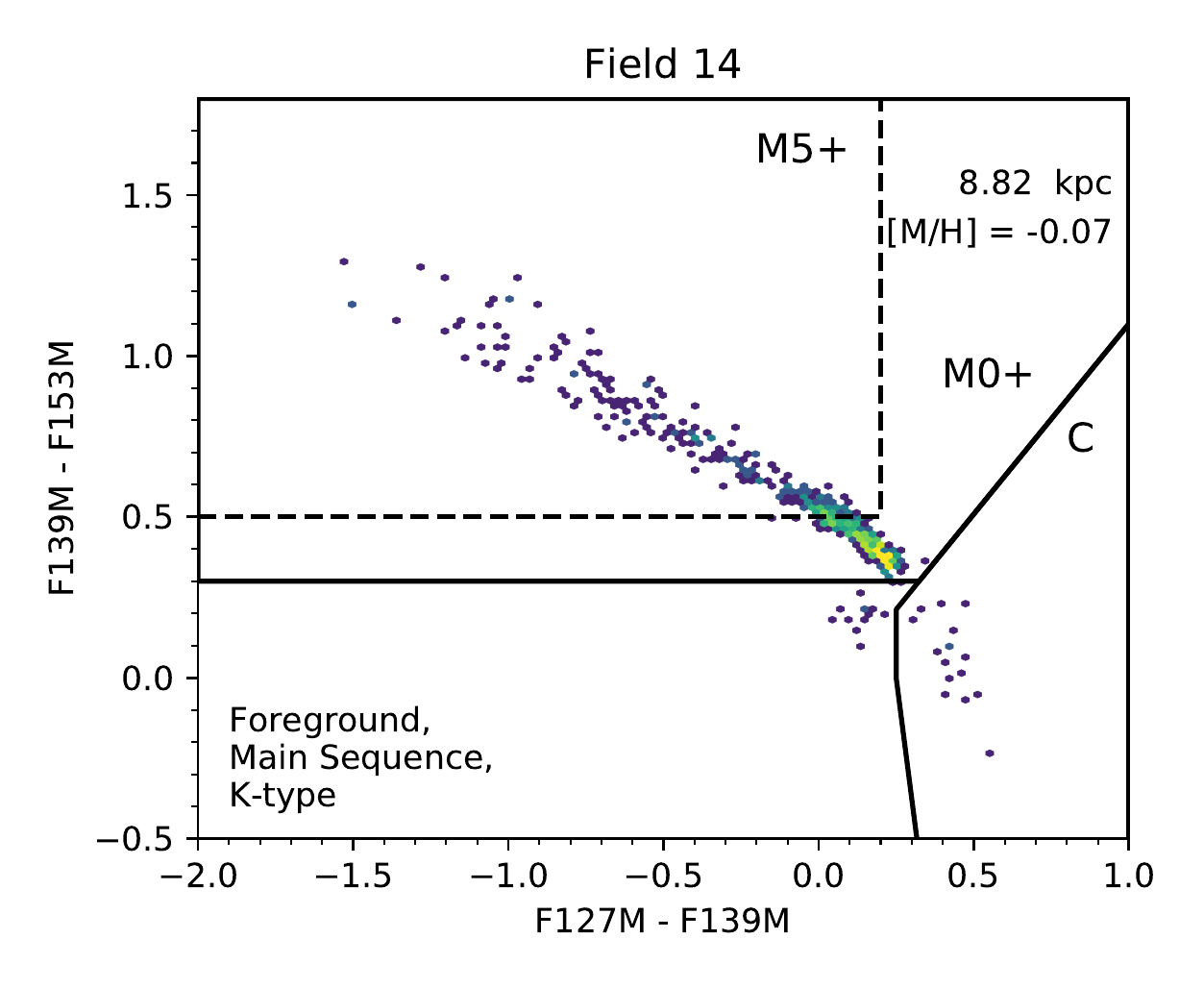}
}                                        
                                         
  \hbox{                                 
  \includegraphics[width=0.31\textwidth]{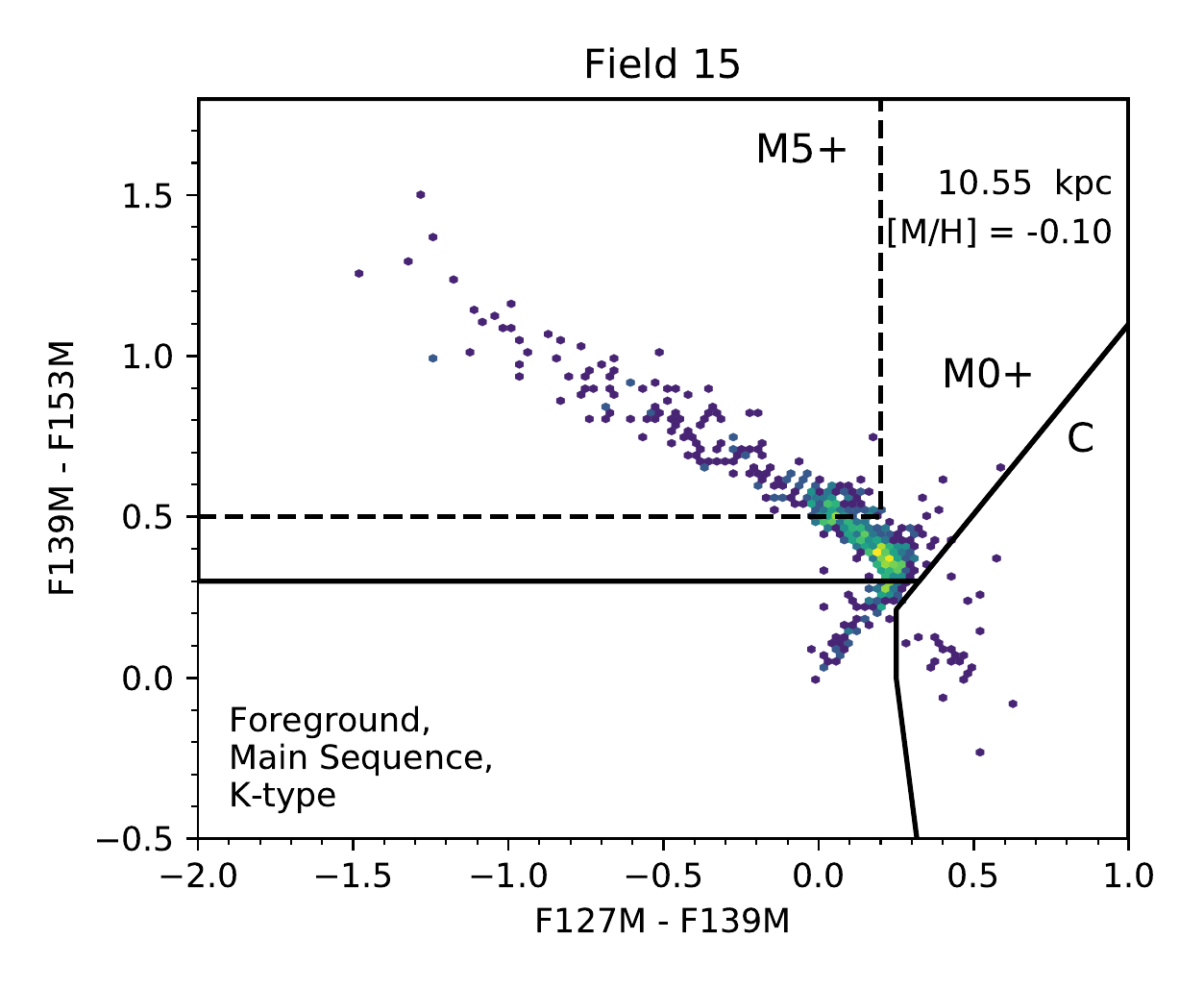}
  \includegraphics[width=0.31\textwidth]{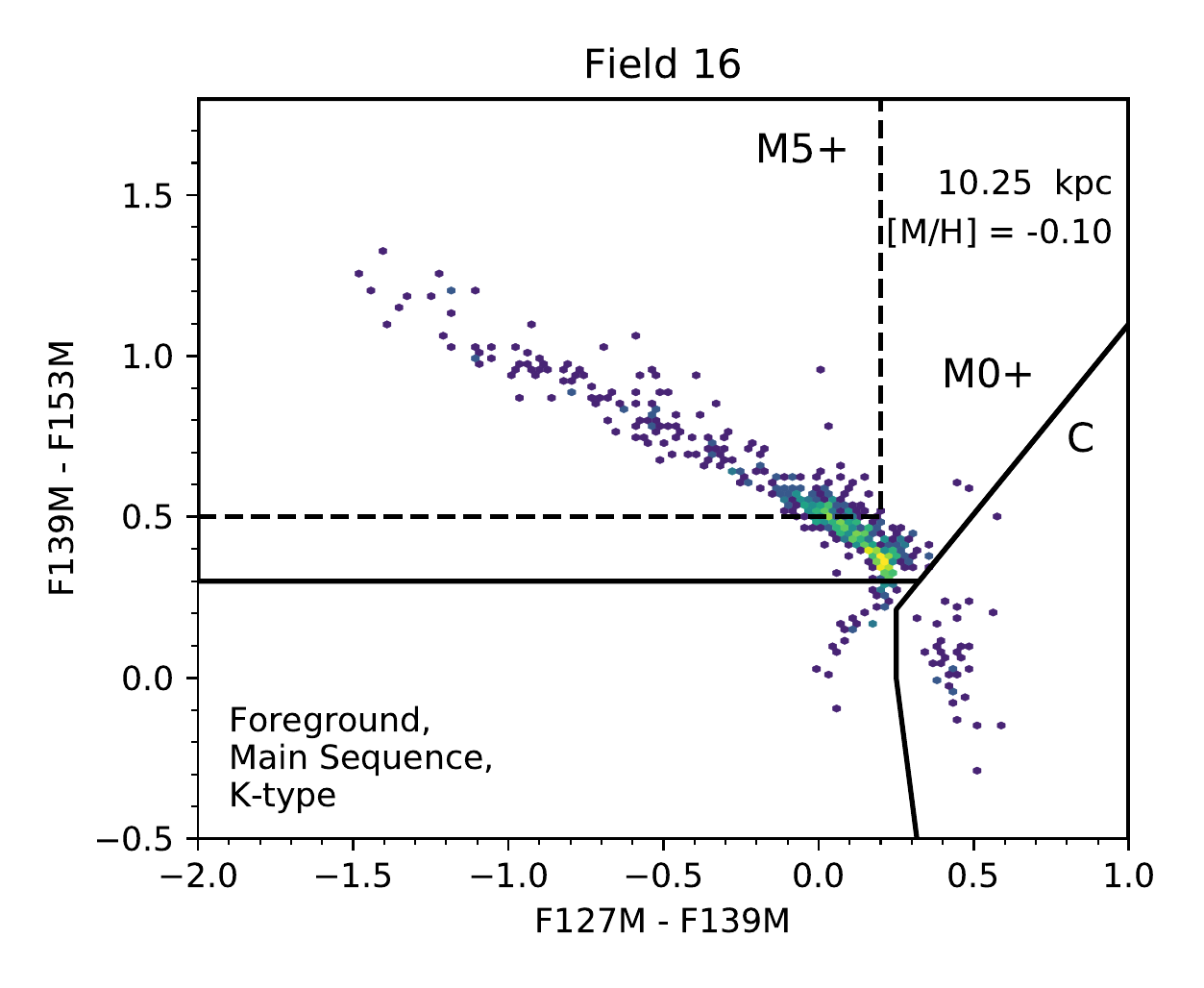}
  \includegraphics[width=0.31\textwidth]{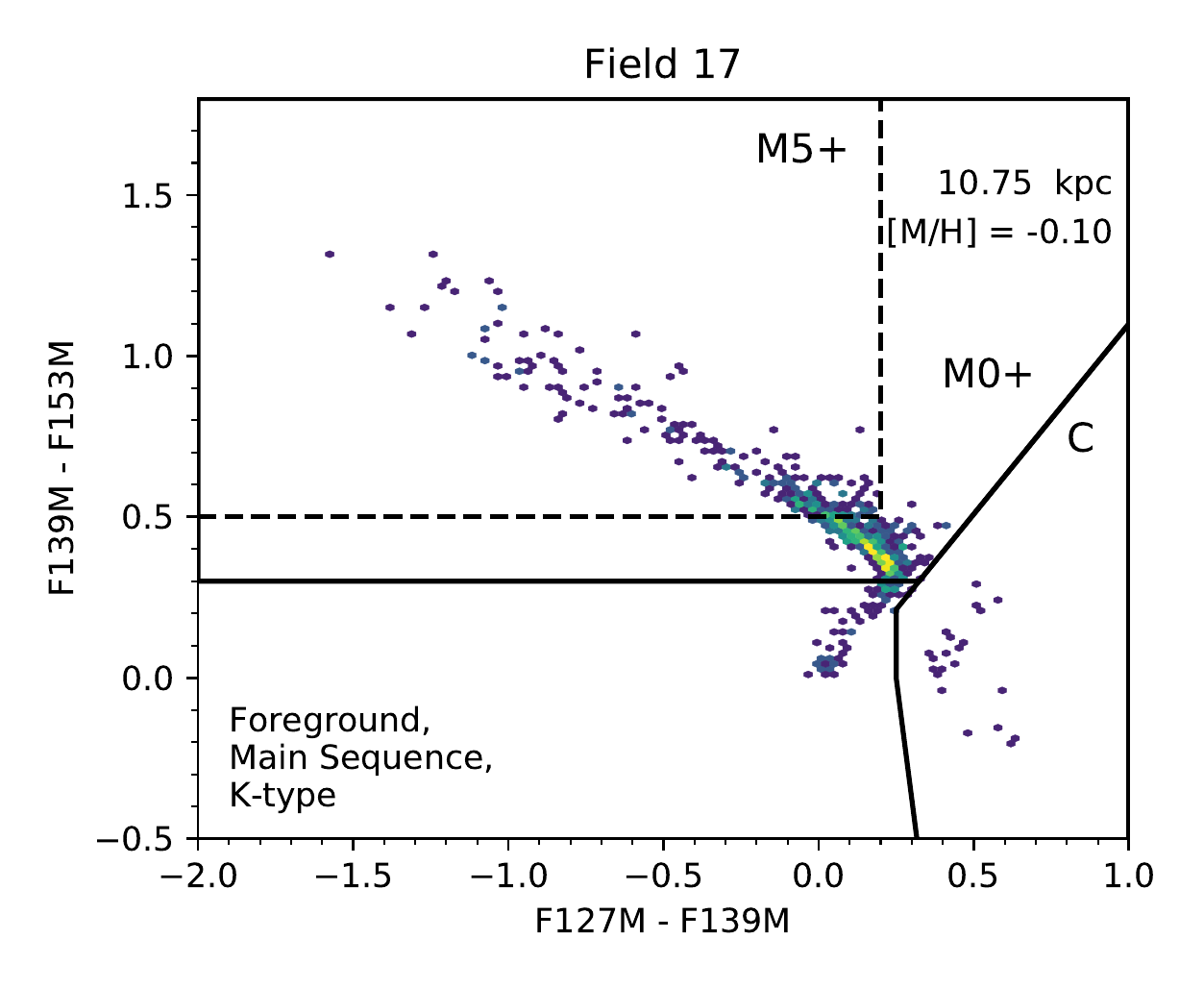}
}                                        
  \hbox{                                 
  \includegraphics[width=0.31\textwidth]{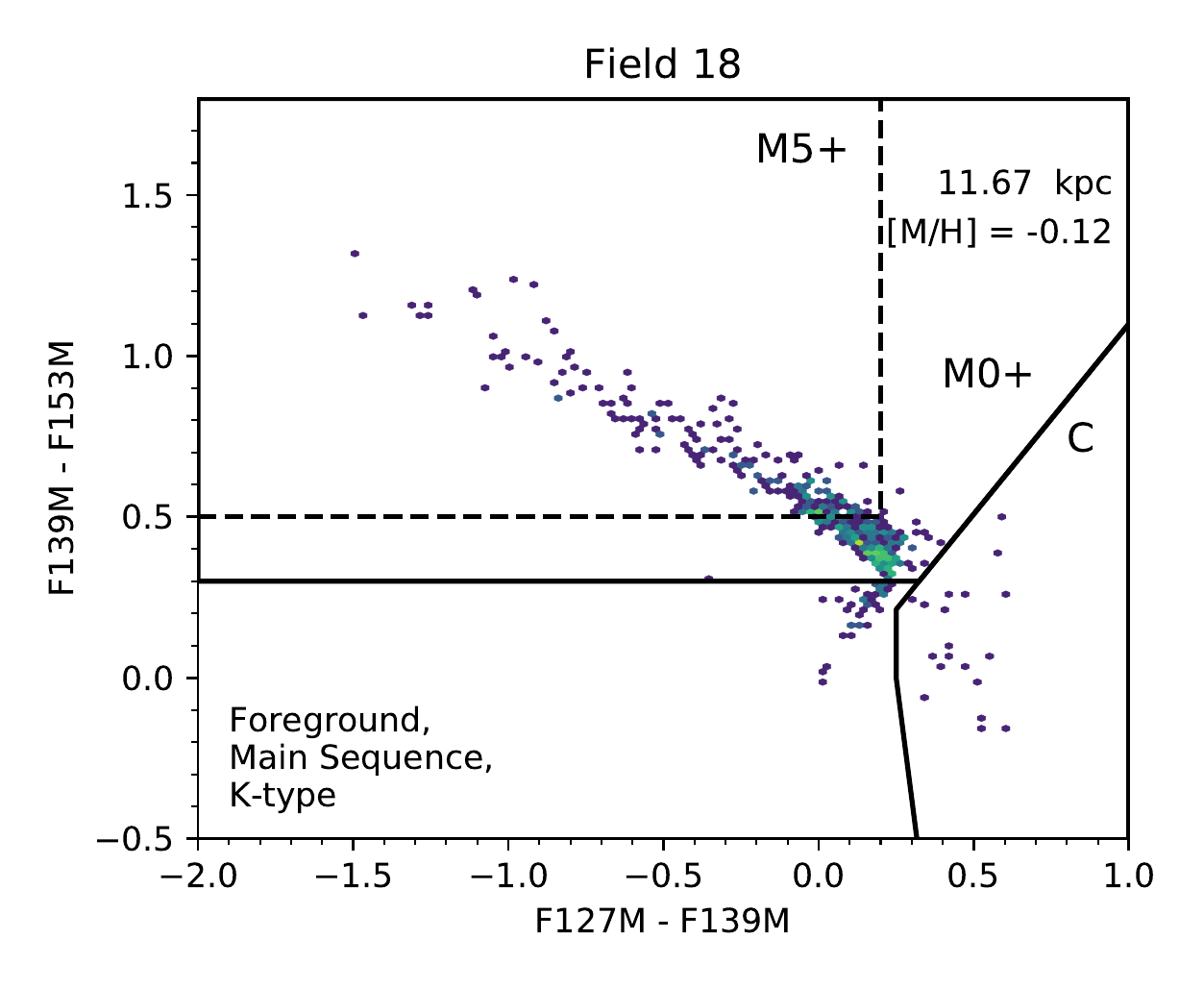}
  \includegraphics[width=0.31\textwidth]{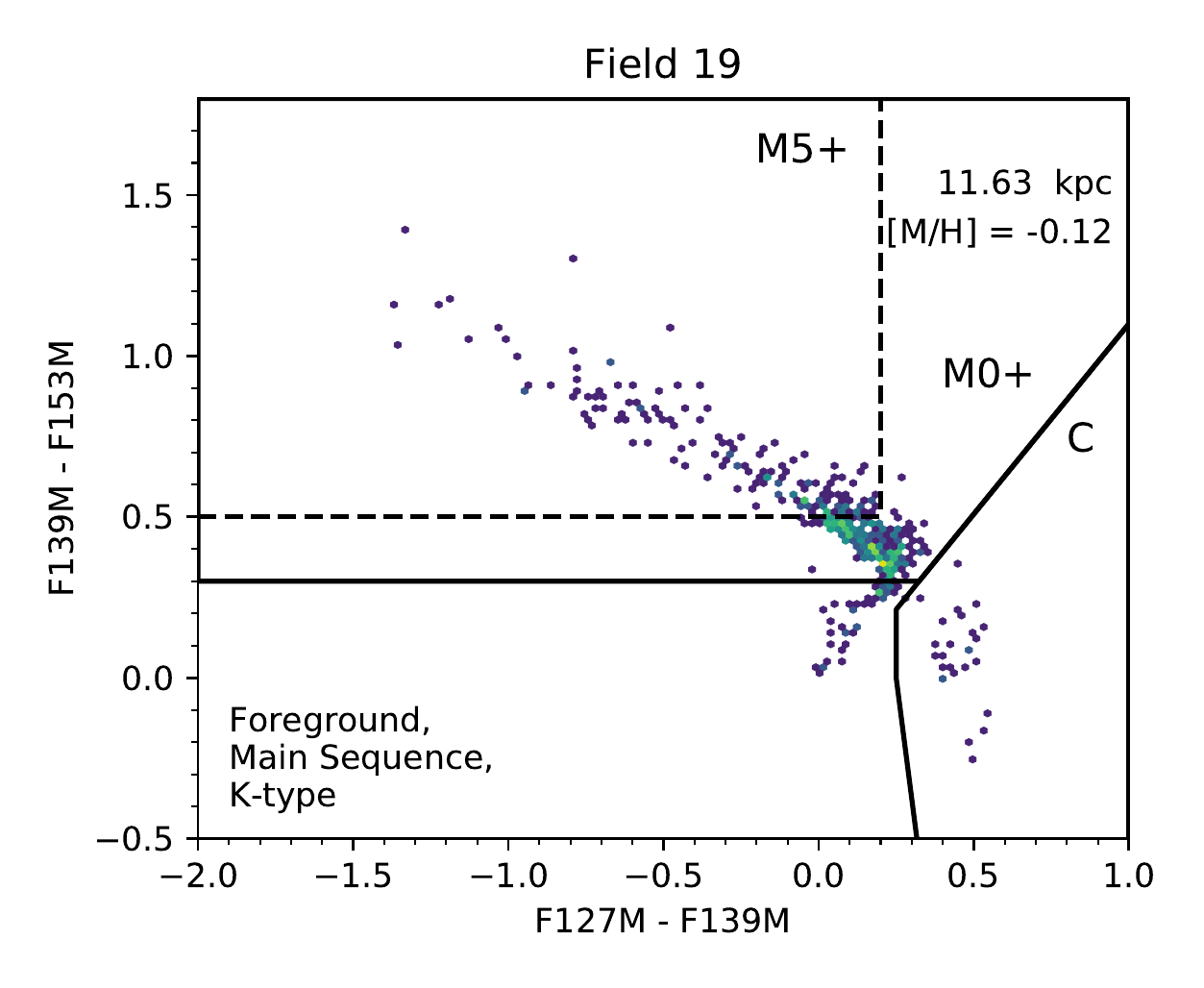}
  \includegraphics[width=0.31\textwidth]{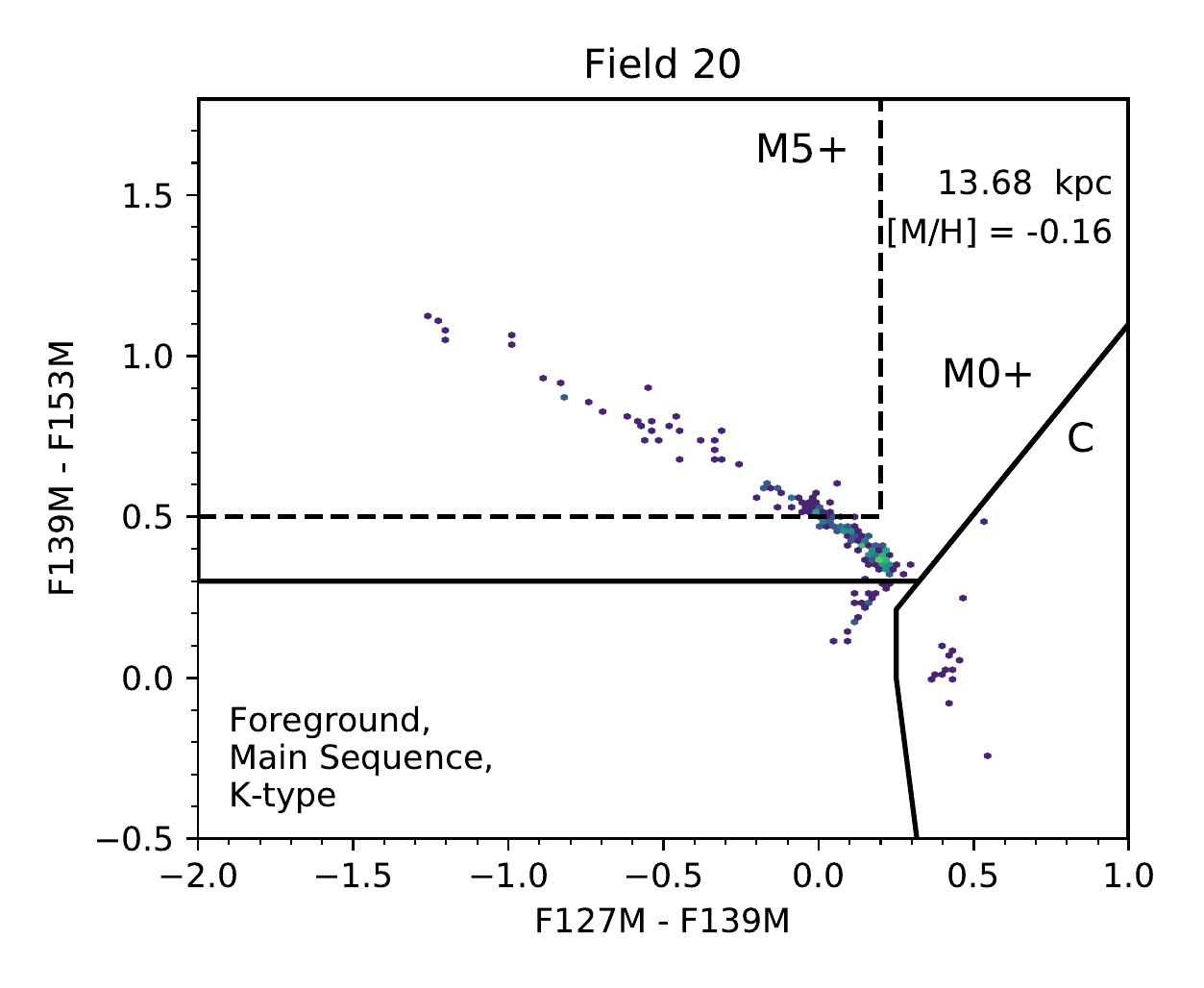}
}
  \caption{Same as Figure~\ref{fig:appendix_ccd1}, for Fields 12-20.}
  \label{fig:appendix_ccd2}
\end{figure}

\end{document}